\documentclass[10pt]{article}
\pdfoutput=1
\usepackage{epsfig}
\usepackage{natbib}

\usepackage{graphicx,graphics}
\usepackage{amssymb,amsmath}
\usepackage{verbatim,enumerate}
\usepackage{rotating}
\usepackage{verbatim,multicol,color}
\usepackage{lmodern}
\usepackage{array}
\usepackage{multirow}
\usepackage{tabularx}
\usepackage{lscape, pdflscape}
\usepackage{url}
\usepackage{dcolumn}
\newcolumntype{d}{D{.}{.}{2.5}}
\newcolumntype{s}{D{.}{.}{1.2}}
\RequirePackage{lineno}

\def\log{\hbox{log}}

\def\boxit#1{\vbox{\hrule\hbox{\vrule\kern6pt
          \vbox{\kern6pt#1\kern6pt}\kern6pt\vrule}\hrule}}

\def\bse{\begin{eqnarray*}}
\def\ese{\end{eqnarray*}}
\def\be{\begin{eqnarray}}
\def\ee{\end{eqnarray}}
\def\bq{\begin{equation}}
\def\eq{\end{equation}}
\def\bse{\begin{eqnarray*}}
\def\ese{\end{eqnarray*}}

\newtheorem{theorem}{Theorem}[section]

\newtheorem{proposition}[theorem]{Proposition}

\newcommand{\qed}{\nobreak \ifvmode \relax \else
      \ifdim\lastskip<1.5em \hskip-\lastskip
      \hskip1.5em plus0em minus0.5em \fi \nobreak
      \vrule height0.75em width0.5em depth0.25em\fi}

\def\part{\partial}

\begin{document}

{\center\LARGE Statistical Analysis of Autoregressive Fractionally Integrated Moving Average Models\\}
\vskip20mm

{\center {\bf Javier E. Contreras-Reyes}\\
\vskip2mm
Instituto de Fomento Pesquero\\

Valpara\'iso, Chile\\

Emails: javier.contreras@ifop.cl, jecontrr@mat.puc.cl\\}
\vskip8mm
{\center {\bf Wilfredo Palma}\\
\vskip2mm
Departamento de Estad\'istica\\

Pontificia Universidad Cat\'olica de Chile\\

Santiago, Chile

Email: wilfredo@mat.puc.cl\\}
\vskip15mm

\begin{abstract}
In practice, several time series exhibit long-range dependence or persistence in their observations, leading to the development of a number of estimation and prediction methodologies to account for the slowly decaying autocorrelations. The autoregressive fractionally integrated moving average (ARFIMA) process is one of the best-known classes of long-memory models. In the package {\tt afmtools} for {\tt R}, we have implemented some of these statistical tools for analyzing ARFIMA models. In particular, this package contains functions for parameter estimation, exact autocovariance calculation, predictive ability testing, and impulse response function, amongst others. Finally, the implemented methods are illustrated with applications to real-life time series.\\

\end{abstract}

\vskip20mm

\section{Introduction}

Long-memory processes introduced by \citet{r:08} and \citet{er:10}, are playing a key role in the time series series literature
(for example see \citet{y:15} and references therein) and have become a useful model for explaining natural events studied in geophysics, biology,
and other areas. As a consequence, a number of techniques for analyzing these processes have been developed and implemented
in statistical packages. For example, packages about long-memory processes have been developed in {\tt R} \citep{R}: the {\tt longmemo}
package produces a Whittle estimation for Fractional Gaussian Noise and Fractional ARIMA models via an approximate
MLE using the \citet{lr:07} algorithm and performs
Spectral Density of Fractional Gaussian Noise and Periodogram Estimate. In addition, the {\tt fracdiff} package simulates ARFIMA
time series, estimates ARFIMA parameters using an approximate MLE approach \citep{b:09},
and calculates their variances with the Hessian method. Recently, \citet{q:11} describe the {\tt forecast} package to automatically predict
univariate time series via State Space models with exponential smoothing for ARIMA models. In addition, the forecast package offers a forecast function
for ARFIMA models estimated using the algorithm proposed by \citet{fr:13}. The afmtools package requires the {\tt polynom}, {\tt hypergeo}, {\tt sandwich} and the aforementioned {\tt fracdiff} and {\tt longmemo} packages.

Unfortunately, many of these computational implementations have important shortcomings. For instance, there is a severe lack of
algorithms for calculating exact autocovariance functions (ACVF) of ARFIMA models, for computing
precise estimator variances, and for forecasting performance tests \citep{m:06}, and impulse
response functions \citep{ha:12}, as well as for other aspects. In order to circumvent some of these problems, this paper discusses the
package {\tt afmtools} developed by \citet{cgp11}. This package
aims to provide functions for computing ACVFs by means of the \citet{v:20} algorithm, ARFIMA
fitting through an approximate estimation scheme via Whittle algorithm \citep{bn:21},  asymptotic
parameter estimate variances and several other tasks mentioned before.
Hence, the aims of this paper are to analyze the {\tt afmtools} package and to illustrate its theoretical and practical performance,
which complements the existing development packages related to ARFIMA models mentioned above. Specifically, we implement our
findings in a meteorological application about tree ring growth.

The remainder of this paper is structured as follows. Section~2 is devoted to describing the ARFIMA processes and their properties.
This section includes an analysis of the  spectral density, autocovariance function, parameter variance-covariance matrix estimation,
impulse response function, and a model parameters estimation method. In addition, this section provides a test for assessing
the predictive ability of a time series model. Finally, Section~3 addresses the performance of the functions implemented
in the  {afmtools} package. Apart from describing the methodologies implemented in this package, we also illustrate
their applications to real-life time series data.

\section[ARFIMA processes]{ARFIMA processes}

Recent statistical literature has been concerned with the study of long-memory models that go
beyond the presence of random walks and unit roots in the univariate time series processes.
The autoregressive fractionally integrated moving-average (ARFIMA) process is a class of long-memory models
(\citet{r:08}; \citet{er:10}), the main objective of which is to explicitly account for persistence to incorporate the
long-term correlations in the data. The general expression for ARFIMA processes $\{y_t\}$ may be defined by
the equation

\begin{equation}\label{arfima}
\Phi(B)y_t=\Theta(B)(1-B)^{-d}\varepsilon_t,
\end{equation}

where $\Phi(B)=1-\phi_1 B-\cdots -\phi_p B^p$ and $\Theta(B)=1+\theta_1 B+\cdots +\theta_q B^q$ are the autoregressive and moving-average operators, respectively; $\Phi(B)$ and $\Theta(B)$ have no common roots, $B$ is the backward shift operator and $(1-B)^{-d}$ is the fractional differencing
operator given by the binomial expansion

\begin{equation}\label{bin}
(1-B)^{-d}=\sum_{j=0}^{\infty}\frac{\Gamma(j+d)}{\Gamma(j+1)\Gamma(d)}B^j=\sum_{j=0}^{\infty} \eta_jB^j,
\end{equation}

for $d\in(-1,1/2)$ and $\{\varepsilon_t\}$ is a white noise sequence with zero mean and innovation variance $\sigma^2$. An asymptotic approximation of

\begin{equation}\label{etae}
\eta_j=\frac{\Gamma(j+d)}{\Gamma(j+1)\Gamma(d)}
\end{equation}

for large $j$ is

\begin{equation}\label{asym}
\eta_j\sim \frac{j^{d-1}}{\Gamma(d)},
\end{equation}

where $\Gamma$ is the usual gamma function.

\begin{theorem}\label{T1}
Consider the ARFIMA process defined by \eqref{arfima} and assume that the polynomials $\Phi(\cdot)$ and $\Theta(\cdot)$ have no common zeros and that $d\in (-1,\frac{1}{2})$. Then,

\begin{description}
  \item[a)] If the zeros of $\Phi(\cdot)$ lie outside the unit circle $\{z:|z|=1\}$, then there is a unique stationary solution of \eqref{arfima} given by $y_t=\sum_{j=-\infty}^{\infty}\psi_j \varepsilon_{t-j}$ where $\psi_j$ are the coefficients of the following polynomial $\psi(z)=(1-z)^{-d}\Theta(z)/\Phi(z)$.
  \item[b)] If the zeros of $\Phi(\cdot)$ lie outside the closed unit disk $\{z:|z|\leq1\}$, then the solution $\{y_t\}$ is \emph{causal}.
  \item[c)] If the zeros of $\Theta(\cdot)$ lie outside the closed unit disk $\{z:|z|\leq1\}$, then the solution $\{y_t\}$ is \emph{invertible}.
\end{description}
\end{theorem}

For a proof of Theorem~\ref{T1}, see e.g. Palma (2007). Recall that, according to the representation theorem of \citet{ki:22}, any
stationary process is the sum of a regular process and a singular process; these two processes are orthogonal and the decomposition
is unique. Thus, a stationary purely nondeterministic process may be expressed as

\begin{equation}\label{wold}
y_t=\psi(B)\varepsilon_t=\sum_{j=0}^{\infty}\psi_j \varepsilon_{t-j},
\end{equation}

The spectral measure of the purely nondeterministic process \eqref{wold} is absolutely continuous with
respect to the Lebesgue measure on $[-\pi,\pi]$, where the spectral density of the process \eqref{arfima} can be written as

\begin{align}
f(\lambda)&=\frac{\sigma^2}{2\pi}|\psi(e^{-i\lambda})|^2\label{sp}\\
        &=\frac{\sigma^2}{2\pi}|1-e^{-i\lambda}|^{-2d}\frac{|\Theta(e^{-i\lambda})|^2}{|\Phi(e^{-i\lambda})|^2}\nonumber\\
          &=\frac{\sigma^2}{2\pi}\left(2\sin \frac{\lambda}{2}\right)^{-2d}\frac{|\Theta(e^{-i\lambda})|^2}{|\Phi(e^{-i\lambda})|^2}.\nonumber
\end{align}

where $i$ denotes the imaginary unit. A special case of ARFIMA models is the fractionally differenced process described by \citet{er:10}, in which the polynomials are $\Phi(B)=\Theta(B)=1$ and the spectral density is given by
$$f(\lambda)=\frac{\sigma^2}{2\pi}|1-e^{-i\lambda}|^{-2d}.$$

\subsection[Whittle estimation]{Whittle estimation}

The methodology to approximate MLE is based on the calculation of the periodogram $I(\lambda)$
by means of the \emph{fast Fourier transform} (FFT); e.g., \citet{f:19}, and the use of the
approximation of the Gaussian log-likelihood function due to \citet{bn:21} and by \citet{bg:01}. So, suppose
that the sample vector $Y=(y_1,y_2,\ldots,y_n)$ is normally distributed with zero mean
and autocovariance given by (\ref{ac}) as

\begin{equation}
\gamma(k-j)=\int_{-\pi}^{\pi}f(\lambda)e^{i\lambda (k-j)}d\lambda,
\end{equation}

where $f(\lambda)$ is defined as in \eqref{sp} and is associated with the parameter set ${\bf \Omega}$
of the ARFIMA model defined in \eqref{arfima}. The log likelihood function of the process $Y$ is given by

\begin{equation}\label{log}
L({\bf\Omega})=-\frac{1}{2n}[\log|{\bf\Delta}|-{\bf Y}^{\top}{\bf\Delta}^{-1} {\bf Y}].
\end{equation}

where ${\bf\Delta}=[\gamma(k-j)]$ with $k,j=1,...,n$. For calculating \eqref{log}, two asymptotic approximations are made for the terms $\log(|{\bf\Delta}|)$ and ${\bf Y}^{\top}{\bf\Delta}^{-1} {\bf Y}$ to obtain

\begin{equation}\label{alog}
L({\bf\Omega})\approx -\frac{1}{4\pi}\left[\int_{-\pi}^{\pi}\log[2\pi f(\lambda)]d\lambda + \int_{-\pi}^{\pi}\frac{I(\lambda)}{f(\lambda)}d\lambda\right],
\end{equation}

as $n\to \infty$, where

\begin{equation}
I(\lambda)=\frac{1}{2\pi n}\left|\sum_{j=1}^{n}y_j e^{i\lambda j}\right|^2,
\end{equation}

is the periodogram indicated before. Thus, a discrete version of \eqref{alog} is actually the Riemann
approximation of the integral and is

\begin{equation}\label{mlh}
L({\bf\Omega})\approx -\frac{1}{2n}\left[\sum_{j=1}^{n}\log f(\lambda_j) + \sum_{j=1}^{n}\frac{I(\lambda_j)}{f(\lambda_j)}\right],
\end{equation}

where $\lambda_j=2\pi j/n$ are the Fourier frequencies. Now, to find the estimator of
the parameter vector ${\bf\Omega}$, we use the minimization of $L({\bf\Omega})$ produced
by the  {nlm} function. This non-linear minimization function carries out a minimization
of $L({\bf\Omega})$ using a Newton-type algorithm. Under regularity
conditions according to Theorem~\ref{T2} (see Section 2.2), the Whittle
estimator $\widehat{{\bf\Omega}}$ that maximizes the log-likelihood function given
in \eqref{mlh} is consistent and distributed normally \citep[e.g.][]{da:01}. The following
figures illustrates the performance of the Whittle estimator for an ARFIMA$(1,d,1)$ model.

\begin{figure}[h!]
    \centering
    \includegraphics[width=4.5cm,height=4.5cm]{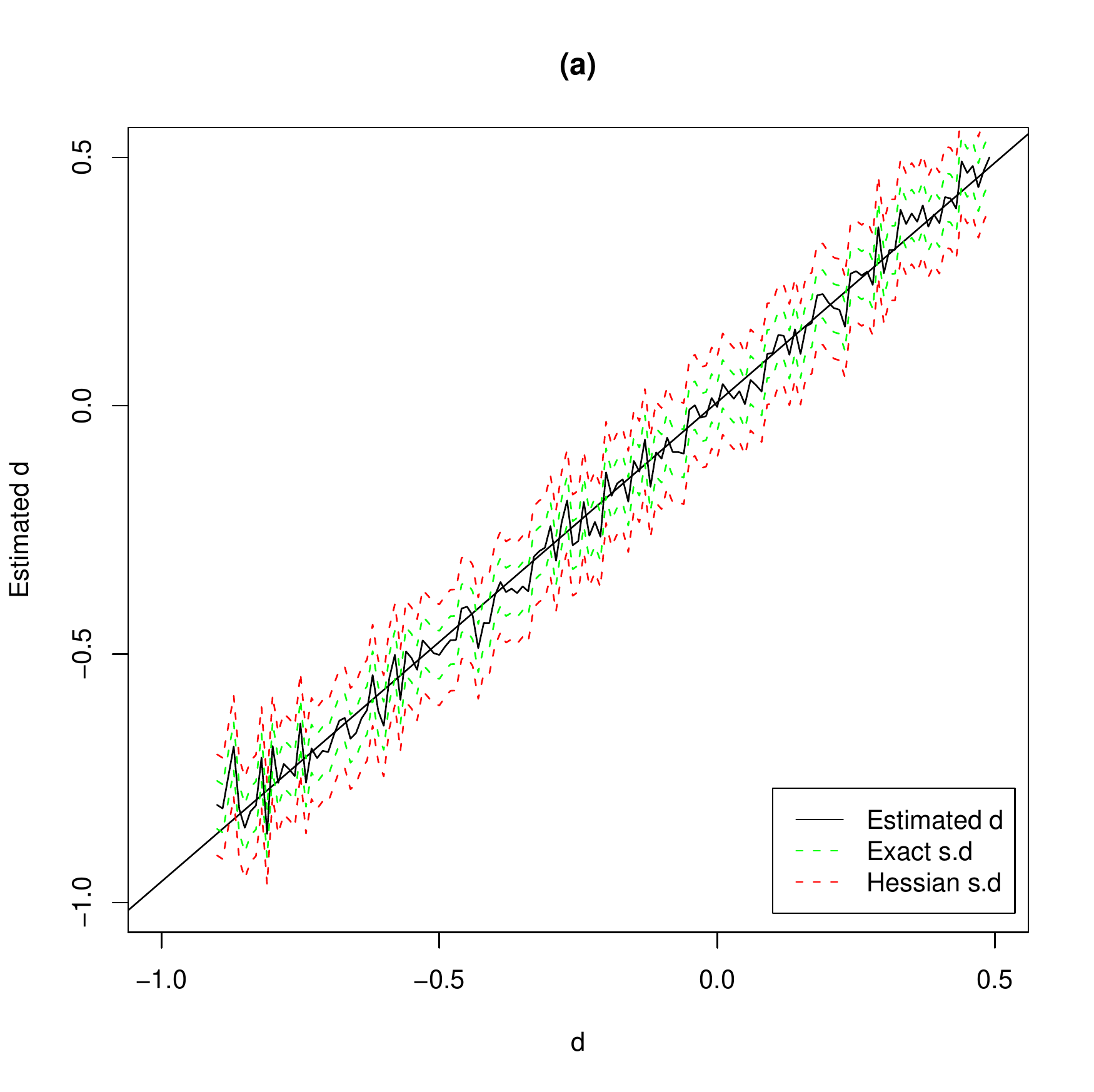}
    \includegraphics[width=4.5cm,height=4.5cm]{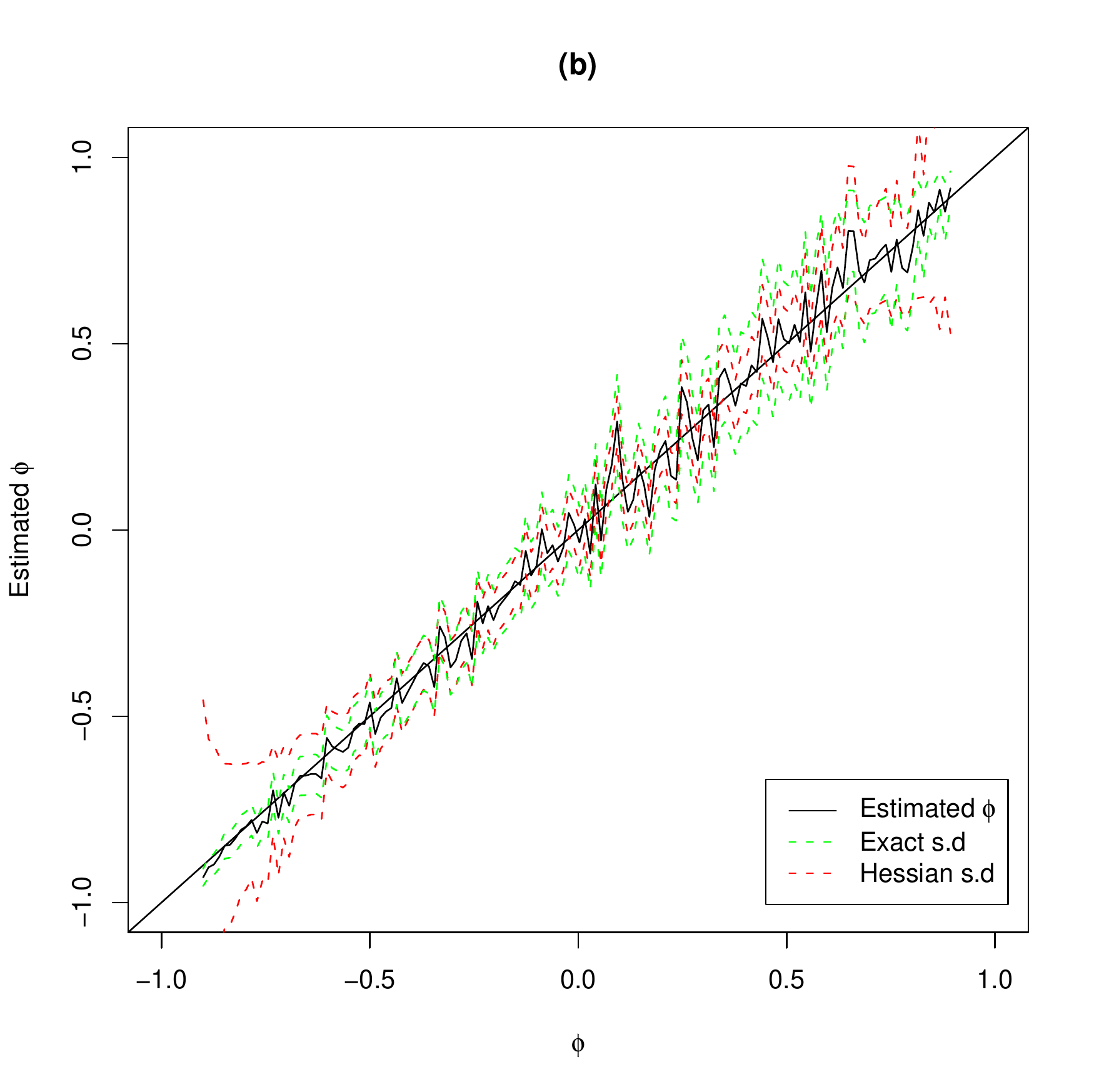}\\
    \includegraphics[width=4.5cm,height=4.5cm]{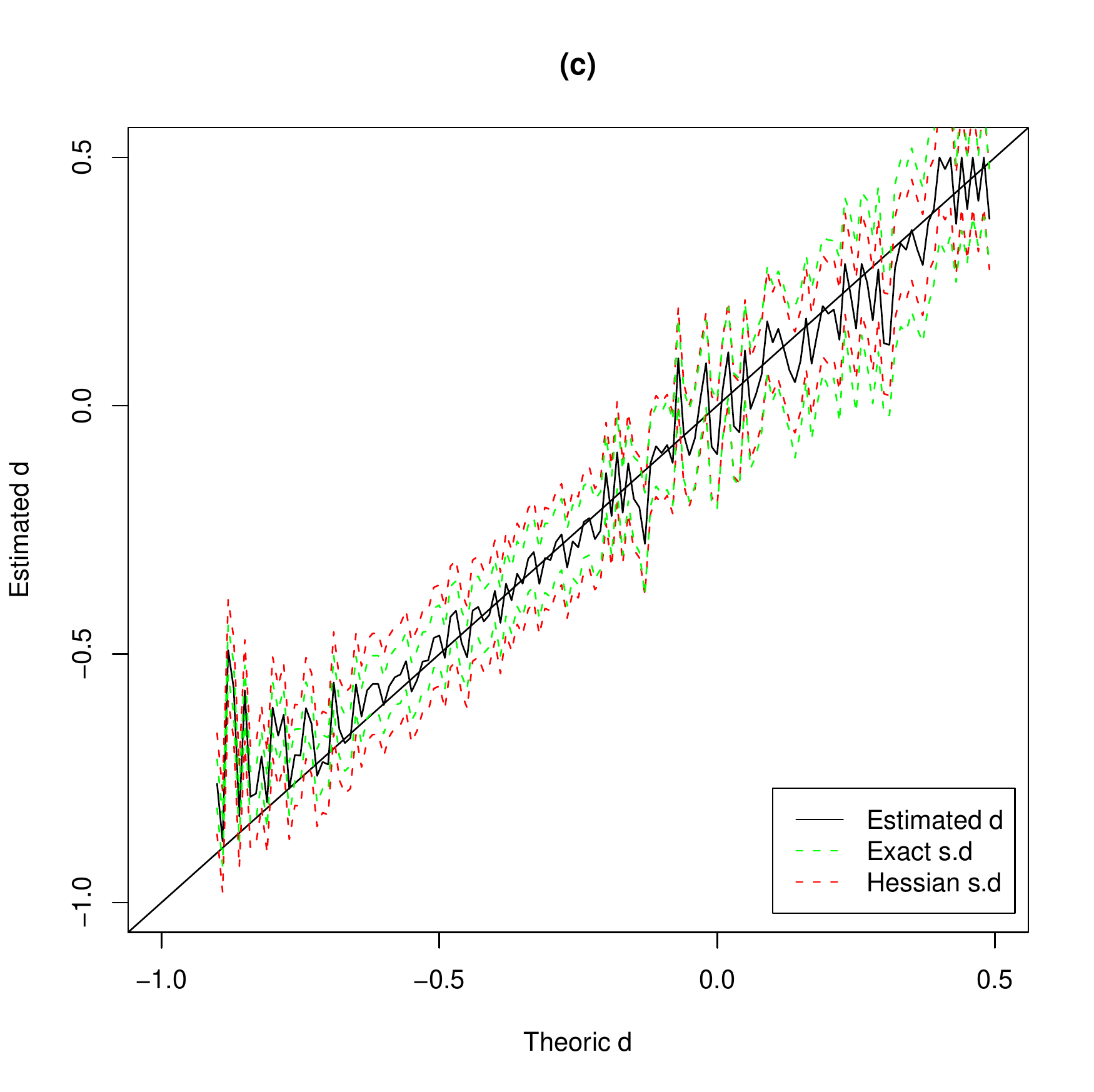}
    \includegraphics[width=4.5cm,height=4.5cm]{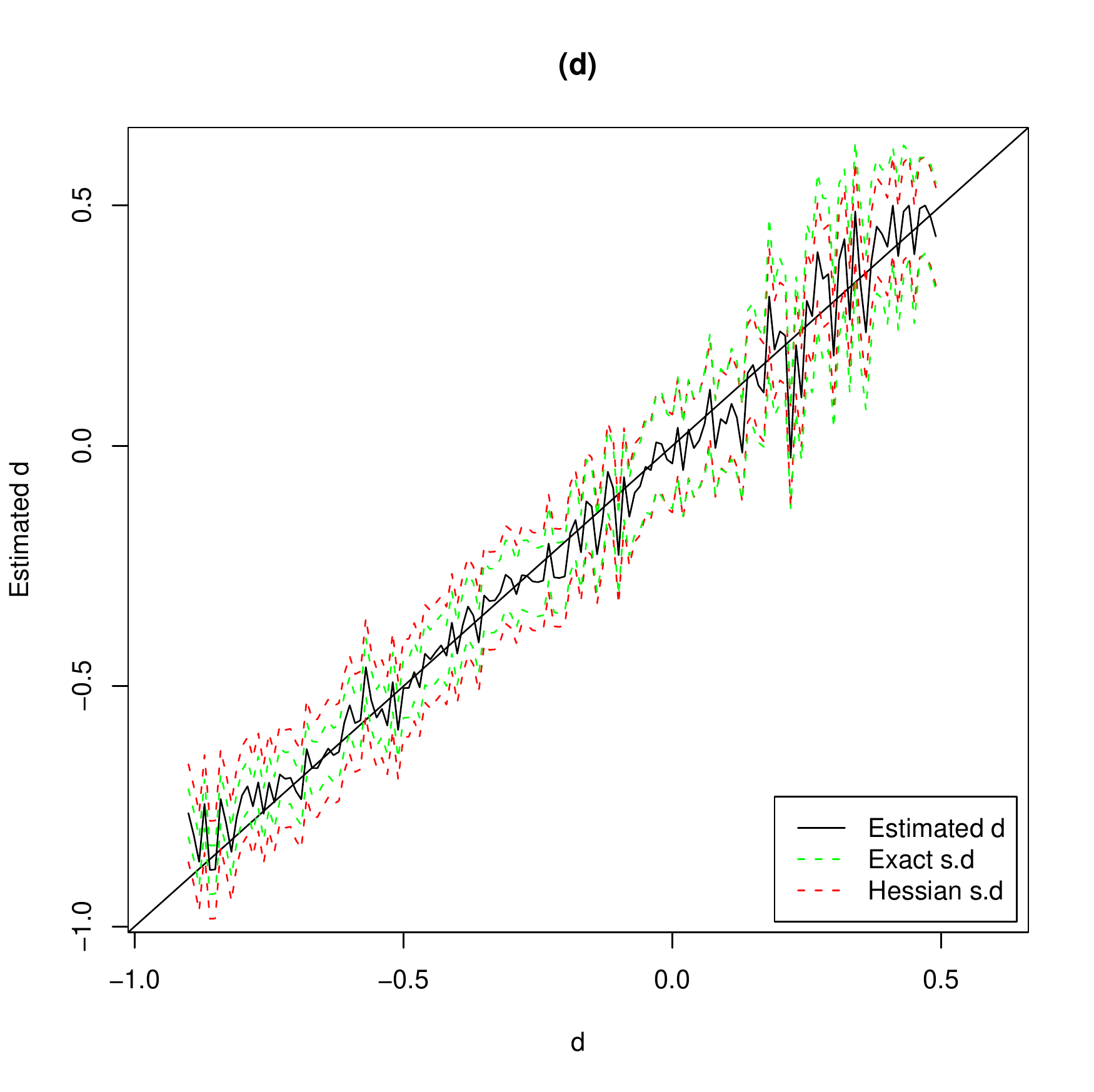}\\
    \includegraphics[width=4.5cm,height=4.5cm]{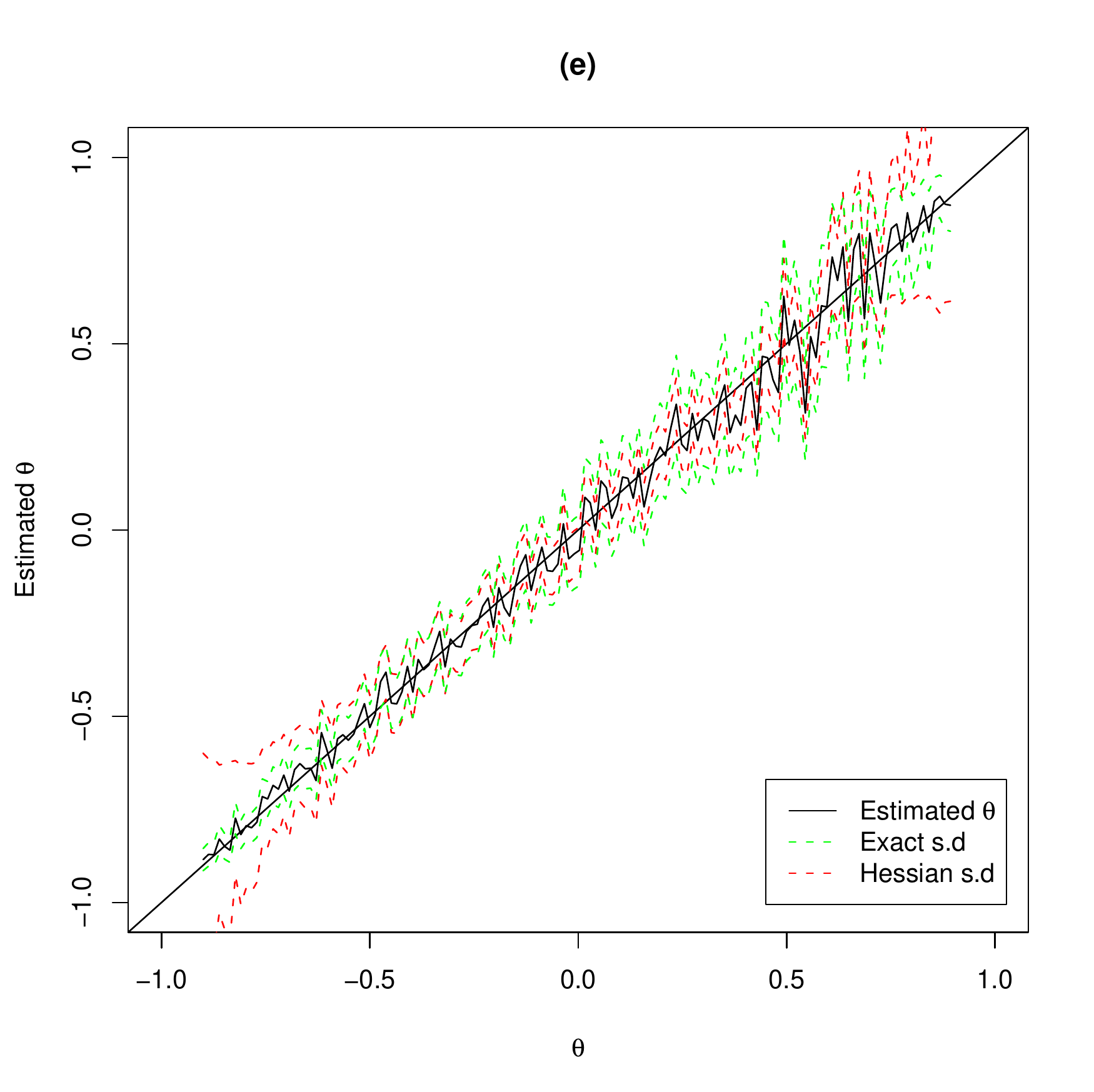}
\caption{Dispersion plots between estimated and theoretical parameters of (a): ARFIMA$(0,d,0)$, (b)-(c): ARFIMA$(1,d,0)$ and (d)-(e): ARFIMA$(0,d,1)$
where the green dotted line is the exact standard deviation  and the red line is the Hessian standard deviation.}\label{sim}
\end{figure}

\begin{figure}[h!]
    \centering
    \includegraphics[width=5cm,height=5cm]{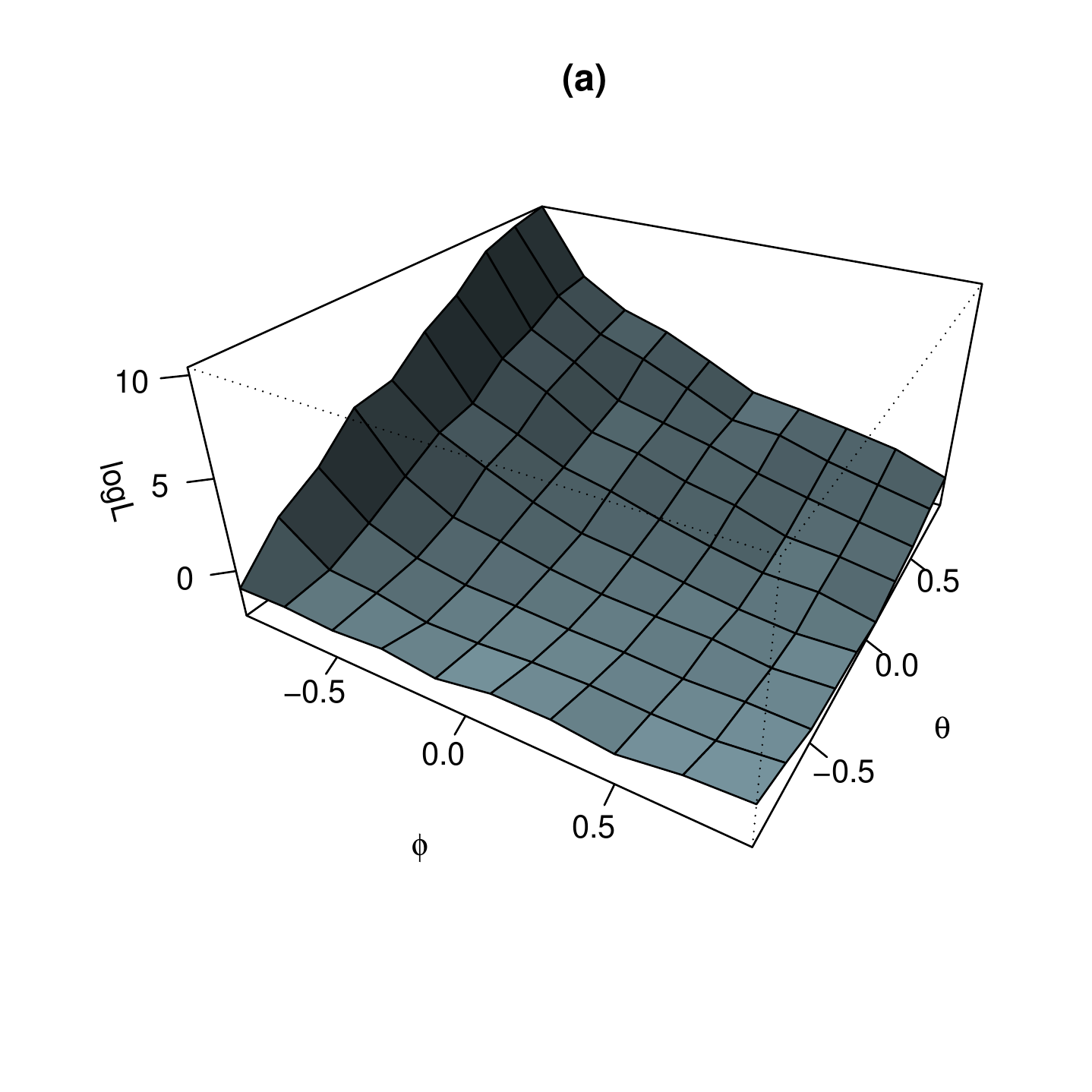}
    \includegraphics[width=5cm,height=5cm]{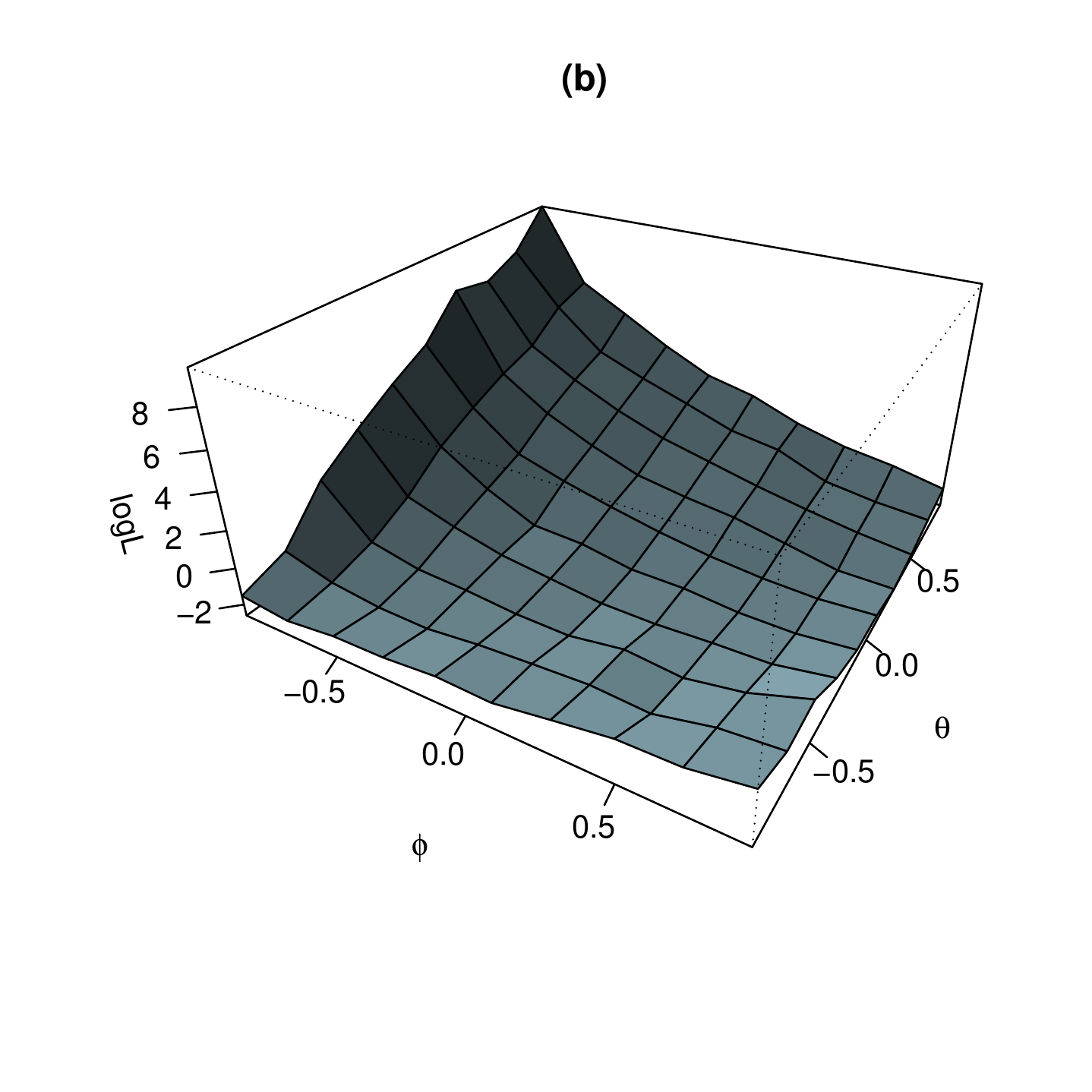}\\
    \includegraphics[width=5cm,height=5cm]{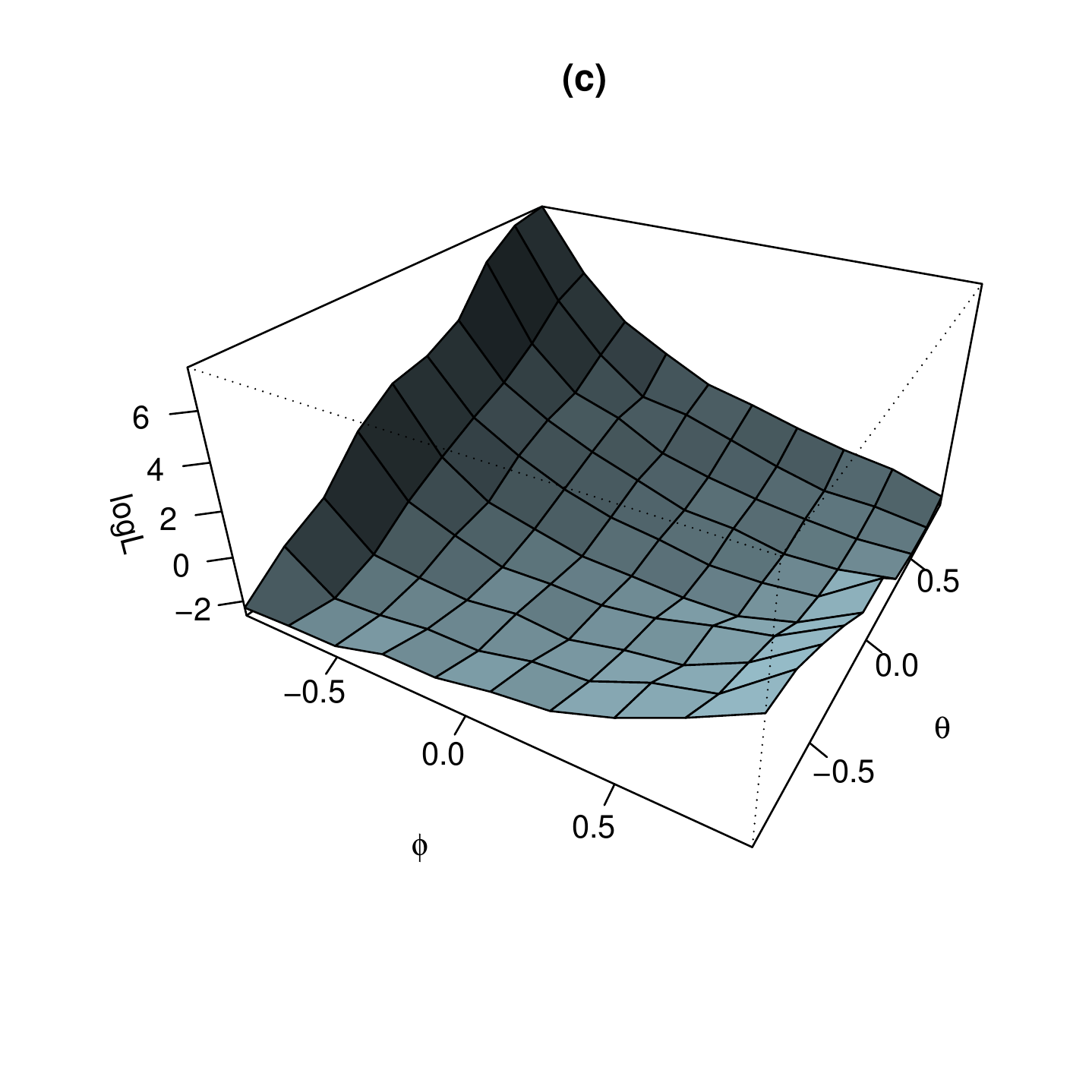}
    \includegraphics[width=5cm,height=5cm]{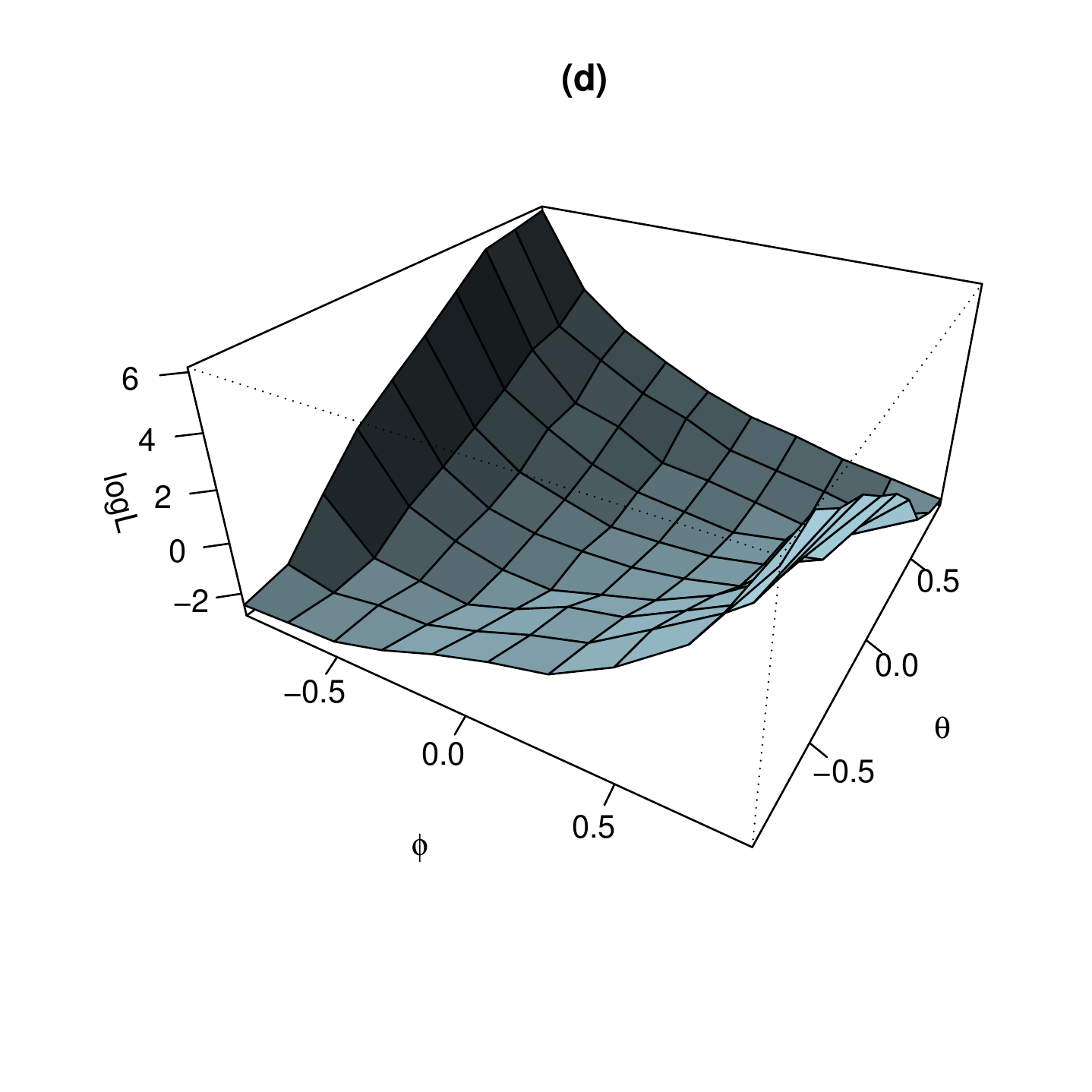}\\
    \includegraphics[width=5cm,height=5cm]{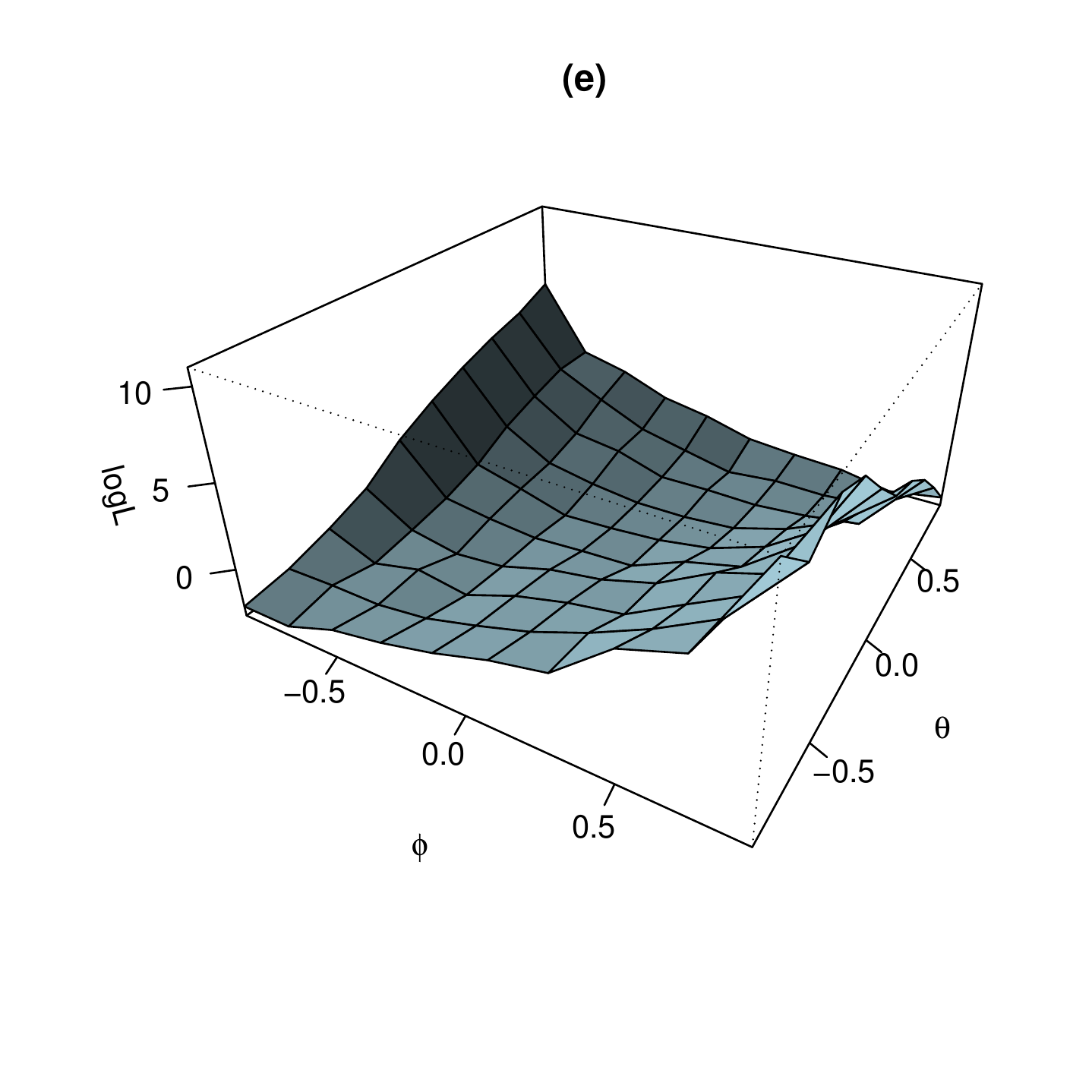}
    \includegraphics[width=5cm,height=5cm]{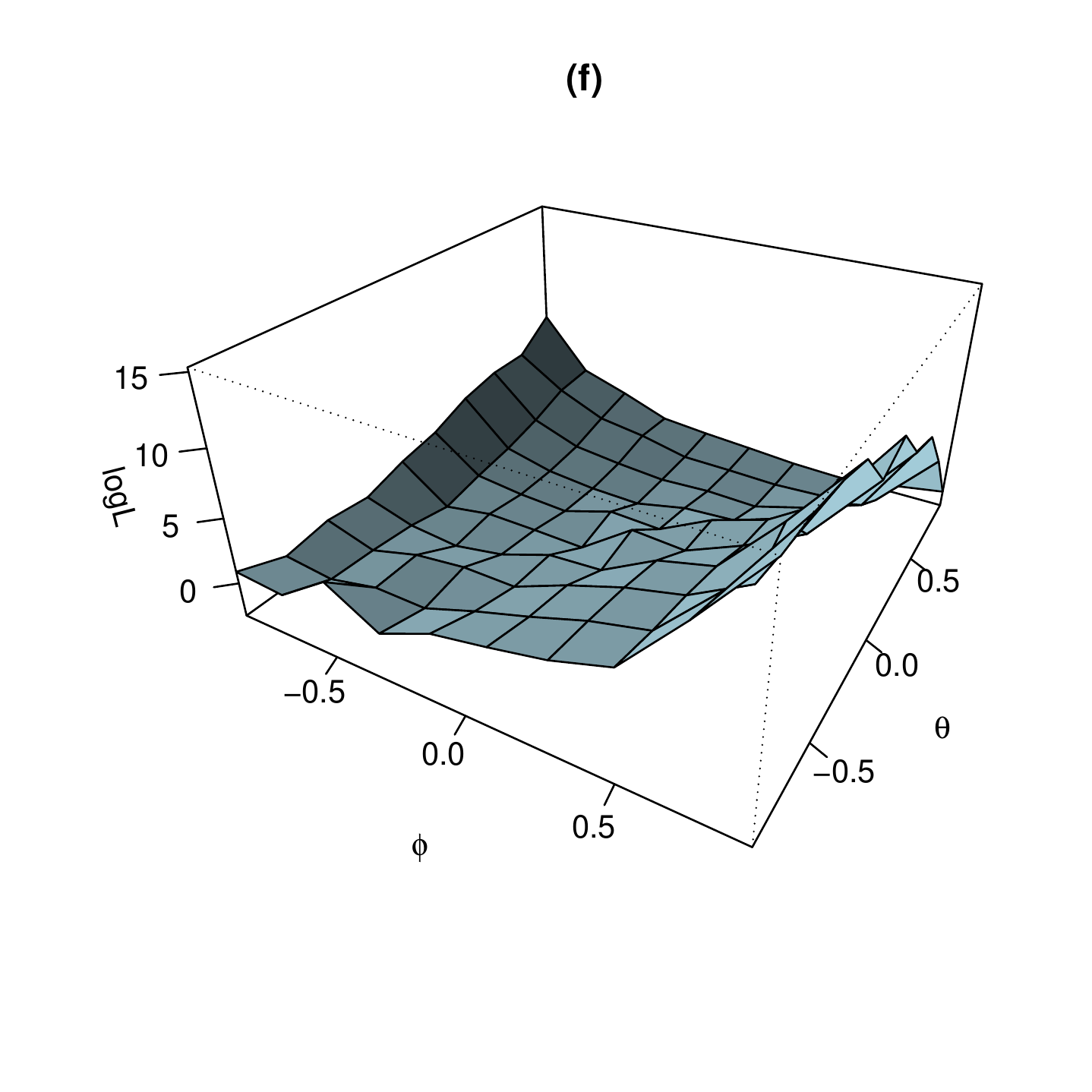}
\caption{Log-Likelihood 3D plots for ARFIMA$(1,d,1)$ model with parameters (a) $d=-0\mbox{.}9$, (b) $d=-0\mbox{.}6$, (c) $d=-0\mbox{.}3$, (d) $d=0$, (e) $d=0\mbox{.}25$ and (f) $d=0\mbox{.}45$
using a grid $\phi\times\theta=(-0\mbox{.}9, 0\mbox{.}9)\times(-0\mbox{.}9, 0\mbox{.}9)$.}\label{likeL}
\end{figure}

Figure~\ref{sim} shows several simulation results assessing the performance of the estimators of $d$, AR, and the MA parameters,
for different ARFIMA models. These plots include the exact and Hessian standard deviations. According to the definition of the ARFIMA model,
the simulations are run in the interval $(-1,0\mbox{.}5)$ for $d$. In addition, Figure~\ref{likeL} shows some simulation
results regarding the log-likelihood behavior for the cases $d=\{-0\mbox{.}9, -0\mbox{.}6, -0\mbox{.}3, 0, 0\mbox{.}25, 0\mbox{.}45\}$
with a rectangular grid $\phi\times\theta=(-0\mbox{.}9, 0\mbox{.}9)\times(-0\mbox{.}9, 0\mbox{.}9)$ for the ARFIMA$(1,d,1)$ model.
The plots of Figure~\ref{sim} show a similar behavior for the estimators with respect to the theoretical parameters,
except for the extreme values of the ARMA parameters near -1 and 1. Consequently, the confidence intervals tend to be
larger than the other values of $\phi$ and $\theta$ parameters for plots (b) and (e). In addition, the plots of
Figure~\ref{likeL} present low values of the likelihood function for the values of $\phi$ and $\theta$ closed to 0,
especially for the plots (c)-(f) when $d=\{-0\mbox{.}3, 0, 0\mbox{.}25, 0\mbox{.}45\}$. However, the plots (a)-(d) show
high values of the likelihood function when this is evaluated for the points near $\phi=-0\mbox{.}9$ and $\theta=0\mbox{.}9$.
For the plots (e)-(f), the behavior is inverse, i.e., the likelihood function tends to be higher for values near
$\phi=0\mbox{.}9$ and $\theta=-0\mbox{.}9$.


\subsection[Parameter variance-covariance matrix]{Parameter variance-covariance matrix}

Here, we discuss a method for calculating the exact asymptotic  variance-covariance
matrix of the parameter estimates. This is a useful tool for making statistical inferences about
exact and approximate maximum likelihood estimators, such as the \citet{b:09} 
and Whittle methods (see Section 2.1). An example of this calculation for an
ARFIMA$(1,d,1)$ model is given by \citet[pp. 105-108]{y:15}. 
This calculation method of the Fisher information matrix is an alternative to the numerical
computation using the Hessian matrix.

This proposed method is based on the explicit formula obtained by means of the  derivatives of the parameters log-likelihood gradients. From the spectral density defined in \eqref{sp}, we define the partial derivatives $\nabla_{\Phi}=(\partial/\partial \phi_i)$ and $\nabla_{\Theta}=(\partial/\partial \theta_j)$, with $i=1,\ldots,p$ and $j=1,\ldots,q$.

\begin{theorem}\label{T2}
Under the assumptions that $y_t$ is a stationary Gaussian sequence, the densities $f(\lambda)$, $f^{-1}(\lambda)$, $\partial/\partial{\bf\mu}_i f^{-1}(\lambda)$, $\partial^2/\partial{\bf\mu}_i\partial{\bf\mu}_jf^{-1}(\lambda)$ and $\partial^3/\partial{\bf\mu}_i\partial{\bf\mu}_j\partial{\bf\mu}_kf^{-1}(\lambda)$ are continuous in $(\lambda,{\bf\mu})$ for a parameter set ${\bf\mu}=\{d,\phi_1,\ldots,\phi_p,\theta_1,\ldots,\theta_q\}$; we have the convergence in distribution for an estimated parameter $\widehat{\bf\mu}$ and the true parameter $\mu_0$ about a Gaussian ARFIMA model with

\begin{equation}
\sqrt{n}(\widehat{\bf\mu}_n - {\bf\mu}_0)\mbox{ }\mathop{\buildrel d\over\longrightarrow}\limits_{n\rightarrow\infty}\mbox{ }
N(0,{\bf\Sigma}^{-1}({\bf\mu}_0)),
\end{equation}

where

\begin{equation} \label{gamma}
{\bf\Sigma}({\bf\mu})=\frac{1}{4\pi}\int_{-\pi}^{\pi}[\nabla \log f_{\bf\mu}(\lambda)][\nabla \log f_{\bf\mu}(\lambda)]^{\top}d\lambda.
\end{equation}
\end{theorem}

For a proof of Theorem~\ref{T2}, see e.g. \citet{y:15}. Thus, if we consider the model \eqref{arfima} with spectral density \eqref{sp} where $\{\varepsilon_t\}$ is an independent and identically distributed $N(0,\sigma^2)$, we have that the parameter variance-covariance matrix $\Gamma$ may be calculated in the following proposition.

\begin{proposition}\label{P1}
If $\{y_t\}$ is stationary, then
\begin{eqnarray*}
\frac{\partial}{\partial d} \log f(\lambda)&=&-\log[2(1-\cos\lambda)],\\
\frac{\partial}{\partial \phi_{\ell}} \log f(\lambda)&=& \frac{\sum_{j=1}^{p}\phi_j \cos[(\ell-j)\lambda]}{\sum_{j=1}^{p}\sum_{k=1}^{p}\phi_j\phi_k \cos[(j-k)\lambda]},\\
\frac{\partial}{\partial \theta_{\ell}} \log f(\lambda)&=&\frac{\sum_{j=1}^{q}\theta_j \cos[(\ell-j)\lambda]}{\sum_{j=1}^{q}\sum_{k=1}^{q}\theta_j\theta_k \cos[(j-k)\lambda]}.
\end{eqnarray*}
\end{proposition}

\paragraph{Proof} First, from the spectral density given in \eqref{sp} we have that

$$\log f(\lambda)=\log\left(\frac{\sigma^2}{2\pi}\right)-d\log[2(1-\cos\lambda)] + \log |\Theta(e^{i\lambda})|^2 - \log |\Phi(e^{i\lambda})|^2.$$

By Theorems~\ref{T1} and \ref{T2}, we observe that $\Phi(e^{i\lambda})=\sum_{j=1}^{p}\phi_j e^{i\lambda j}$, this yields
\begin{eqnarray*}
|\Phi(e^{i\lambda})|^2&=&\sum_{j=1}^{p}\sum_{k=1}^{p}\phi_j\phi_k e^{i\lambda(j-k)}\\
&=&2\sum_{j=1}^{p}\sum_{k=1}^{p}\phi_j\phi_k \cos[(j-k)\lambda],
\end{eqnarray*}
and
\begin{eqnarray*}
\frac{\partial}{\partial \phi_{\ell}}|\Phi(e^{i\lambda})|^2&=&2\phi_{\ell}+\sum_{j\neq\ell}\phi_j e^{i\lambda(\ell-j)} + \sum_{k\neq\ell}\phi_k e^{i\lambda(\ell-k)}\\
&=&\sum_{j=1}^{p}\phi_j e^{i\lambda(\ell-j)} + \sum_{k=1}^{p}\phi_k e^{i\lambda(\ell-k)}\\
&=&2\sum_{j=1}^{p}\phi_j \cos[(\ell-j)\lambda].
\end{eqnarray*}

Analogously, we have that

$$\frac{\partial}{\partial \theta_{\ell}}|\Theta(e^{i\lambda})|^2=2\sum_{j=1}^{q}\theta_j \cos[(\ell-j)\lambda].$$

Then, this implies the results for $\frac{\partial}{\partial \phi_{\ell}} \log f(\lambda)$ and $\frac{\partial}{\partial \theta_{\ell}} \log f(\lambda)$.
For $\frac{\partial}{\partial d} \log f(\lambda)$ is direct.\\

Some computations and implementation of this matrix are described in Section~3.2, associated with the parameters of several ARFIMA models,
using the Whittle estimator.

\subsection[Impulse response functions]{Impulse response functions}

The impulse response functions (IRF) is the most commonly used tool to evaluate the effect of shocks on time series.
Among the several approximations to compute this, we consider the theory proposed by \citet{ha:12} to find
the IRF of a process $\{y_t\}$ following an ARFIMA($p,d,q$) model. The properties of these approximations,
depend on whether the series are assumed to be stationary according to Theorem~\ref{T1}. So, under the assumption that the roots of
the polynomials $\Phi(B)$ and $\Theta(B)$ are outside the closed unit disk and $d\in(-1,1/2)$, the
process $\{y_t\}$ is stationary, causal and invertible. In this case, we can write
$y_t=\Psi(B)\varepsilon_t$ where $\Psi(B)$ represents the expansion of the MA$(\infty)$
coefficients denoted as $\psi_j$ with $j>1$. These coefficients satisfy the asymptotic relationship
$\psi_j\sim [\Theta(1)j^{d-1}]/[\Phi(1)\Gamma(d)]$ as $j\rightarrow\infty$ \citep{w:12},
$\Theta(1)=1+\sum_{i=1}^{q}\theta_i$, and
$\Phi(1)=1-\sum_{i=1}^{p}\phi_i$. As a particular case, we have that the $\psi_j$ coefficients for
an ARFIMA$(0,d,0)$  are given in closed form by the expression
$\psi_j=\Gamma(j+d)/(\Gamma(j+1)\Gamma(d))$. Now, from (\ref{bin}) and the Wold expansion
(\ref{wold}), the process (\ref{arfima}) has the expansion
$(1-B)^{-d}y_t=\sum_{j=0}^{\infty}\eta_jy_{t-j}=\sum_{j=0}^{\infty}R_j\varepsilon_{t-j}$, where
$R_j$ is the so-called IRF and is given by

\begin{equation}\label{IR1}
R_j=\sum_{i=0}^{j}\psi_i\eta_{j-i}.
\end{equation}

The terms $\eta_j$ can be represented in recursive form using (\ref{etae}) as $\eta_j=\left(1+(d-1)/j\right)\eta_{j-1}$,
for $j\geq 1$ and $\eta_0=1$. From the asymptotic expression given in (\ref{asym}) and assuming that $\sum_{j=0}^{\infty}\psi_j<\infty$,
we have the following asymptotic representation

\begin{equation}\label{IR2}
R_j\sim\frac{j^{d-1}}{\Gamma(d)}\sum_{i=0}^{\infty}\psi_i
\end{equation}

as $j\rightarrow\infty$ and $\psi_j/(j^{d-1})\longrightarrow0$.\\ 

\subsection[Autocovariance function]{Autocovariance function}

We illustrate a method to compute the exact autocovariance function for the general ARFIMA$(p,d,q)$
process. Considering the parameterization of the autocovariance function derived by writing the spectral density (\ref{sp}) in
terms of parameters of the model given by \citet{v:20},  the autocovariance function of a general
ARFIMA$(p,d,q)$ process is given by

\begin{equation}\label{ac}
\gamma(h)=\frac{1}{2\pi}\int_{0}^{2\pi}f(\lambda)e^{-i\lambda h}d\lambda ,
\end{equation}

where $i$ denotes the imaginary unit. Particularly, the autocovariance and autocorrelation functions of the fractionally
differenced ARFIMA$(0,d,0)$ process are given by

\begin{eqnarray*}
\gamma_0(h)&=&\sigma^2 \frac{\Gamma(1-2d)}{\Gamma(1-d)\Gamma(d)}\frac{\Gamma(h+d)}{\Gamma(1+h-d)},\\
\rho_0(h)&=&\frac{\Gamma(1-d)}{\Gamma(d)}\frac{\Gamma(h+d)}{\Gamma(1+h-d)},\\
\end{eqnarray*}

respectively. Then, the polynomial $\Phi(B)$ in \eqref{arfima} may be written as

\begin{equation}\label{rhok}
\Phi(B)=\prod_{i=1}^{p}(1-\rho_i B).
\end{equation}

Under the assumption that all the roots of $\phi(B)$ have multiplicity one, it can be deduced from \eqref{ac} that

$$\gamma(h)=\sigma^2 \sum_{i=-q}^{q} \sum_{j=1}^{p} \psi(i)\xi_j C(d,p+i-h,\rho_j).$$

with

\begin{eqnarray*}
\xi_j&=&\left[\rho_j\prod_{i=1}^{p}(1-\rho_i\rho_j)\prod_{k\neq j}(\rho_j-\rho_k)\right]^{-1},\\
C(d,h,\rho)&=&\frac{\gamma_0(h)}{\sigma^2}[\rho^{2p}\beta(h)+\beta(-h)-1],\\
\beta(h)&=&F(d+h,1,1-d+h,\rho),\\
F(a,b,c,x)&=&1+\frac{a\cdot b}{c\cdot 1}x + \frac{a\cdot (a+1)\cdot b\cdot (b+1)}{c\cdot (c + 1)\cdot 1\cdot 2}x^2 + \ldots
\end{eqnarray*}

where $F(a,b,c,x)$ is the Gaussian hypergeometric function \citep[e.g.][]{mn:07}.
The term $\psi(i)$ presented here and in \citet[pp. 47-48]{y:15} is a
corrected version with respect to \citet{v:20} and is

$$\psi(i)=\sum_{k=\max(0,i)}^{\min(q,q+i)}\theta_k\theta_{k-i}.$$

In the absence of AR parameters the formula for $\gamma(h)$ reduces to

$$\gamma(h)=\sigma^2 \sum_{i=-q}^{q} \psi(i) \frac{\Gamma(1-2d)\Gamma(h+d-i)}{\Gamma(1-d)\Gamma(d)\Gamma(1+i-d-h)}.$$

On the other hand, the findings of \citet{ha:12} describe the asymptotic
behavior of the autocovariance function $\gamma(h)$ as

\begin{equation}\label{ACVFa}
\gamma(h)\sim c_{\gamma}|h|^{2d-1},
\end{equation}

where $c_{\gamma}=\sigma^{2}\pi^{-1}\Gamma(1-2d)\sin(\pi d)\left(\sum_{j=0}^{\infty}\psi_j\right)^2$ for large $|h|$. Let $\{y_1,y_2,\ldots,y_n\}$ be a sample from the process in \eqref{arfima} and let $\overline{y}$ be the sample mean. The exact variance of $\overline{y}$ is given by

\begin{eqnarray*}
\mbox{Var}(\overline{y})=\frac{1}{n}\left[2
\sum_{j=1}^{n-1}\left(1-\frac{j}{n}\right)\gamma(j)+\gamma(0)\right].
\end{eqnarray*}

By \eqref{ACVFa} and for large $n$, we have the asymptotic variance formula $\mbox{Var}(\overline{y})\sim n^{2d-1}c_{\gamma}/d(2d+1)$. The Sowell method is implemented in Section~3.3 for a selected ARFIMA$(1,d,1)$ model
and its exact autocorrelation function is  compared with the sample autocorrelation.\\

\subsection[Predictive ability test]{Predictive ability test}
One approach to compare prediction models is through their root mean square  error (RMSE). Under
this paradigm and given two forecasting methods, the one that presents the lower RMSE is the better of the two. To
compare statistically the differences of predictive ability among two proposed models, we focus here
on the evaluation paradigm proposed by \citet[GW]{m:06}. This test aims to evaluate a
prediction method but not to carry out a diagnostic analysis. Therefore, it does not consider the
parametric uncertainty, which is useful if we want to compare nested models from an ARFIMA model.
The GW test attributed to \citet{llt:05} is based on the differences
$\Delta L_i=|\widehat{x}_i - y_i|-|\widehat{z}_i - y_i|$,
where $\widehat{x}_i$ and $\widehat{z}_i$ are the forecasted observations of the first and second model respectively, for $i=1,...,n$.
The null hypothesis for the GW test associated with expected difference $E[\Delta L]$ of the two prediction models is $H_0$:
$E[\Delta L]=0$, whereas the alternative is $H_1$: $E[\Delta L]\neq 0$. These hypotheses are tested by
means of the  statistic,
$$\Delta\widehat{L}(N)=\frac{1}{N\sqrt{\frac{\widehat{\sigma}^2_N}{N}}}\sum_{i=t_0}^{n-\tau}\Delta L_i,$$
where $N=n-\tau-t_0+1$, $n$ is the total size of the sample, $\tau$ is the prediction horizon, and $t_0$ is the observation at which the mobile windows start.
Note that under $H_0$, the statistic $\Delta \widehat{L}(N)$ is asymptotically normal. For $\tau=1$,
an estimator of $\widehat{\sigma}_{N}$ can be obtained  from the estimation of $\widehat{\sigma}_{\widehat{\alpha}}$ from a
simple regression $\Delta \widehat{L}(N)=\widehat{\alpha} + \epsilon$. However, for horizons $\tau>1$, it is possible to apply a
heteroscedasticity and autocorrelation consistent (HAC) estimator; for example, \citet{z:14} or \citet{bs:01}. The GW test is
implemented in the {\tt gw.test()} function and described in Section~3.4.

\section[Application]{Application}

\begin{figure}[h!]
    \centering
    \includegraphics[width=6.5cm,height=3.5cm]{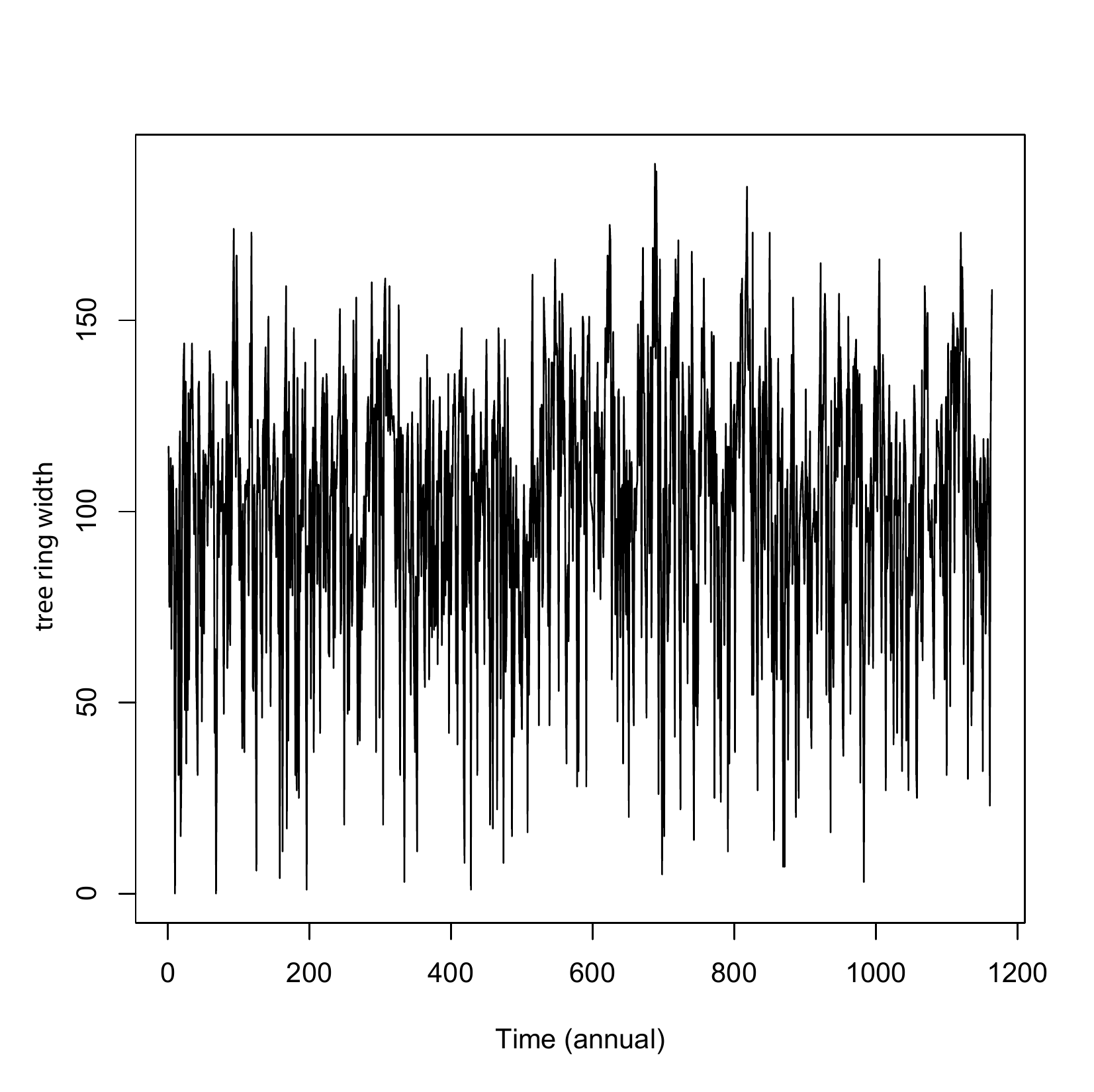}
    \includegraphics[width=4.5cm,height=3cm]{tree_rings-v2.jpg}
\caption{Left: Tree Ring Data Base. Right: Illustration of the observed climatic episodes and tree ring growth. Image source: http://summitcountyvoice.com}\label{tree}
\end{figure}

The functions described in the previous sections are implemented in the {\tt afmtools} package. We illustrate the performance of the Whittle method by applications to real-life time series {\tt TreeRing} (Statlib Data Base, http://lib.stat.cmu.edu/) displayed in Figure~\ref{tree} left. In climatological areas, it is very important to analyze this kind of data because this allows us to explore rainy and dry seasons in the study area. \citet{w:15} has been modeling this time series to determine the range of possible growths for the upcoming years of the trees using ARMA and ARIMA models. On the other hand, this time series displays a high persistence in its observations and has been analyzed by \citet{h:16} and \citet{pof:16} with a locally stationary approach. Apparently, the illustrated growth of the trees, represented by the number of the rings, presents a long-range dependence of its observations along the observations for ages and seasons (see Figure~\ref{tree}, right). For these reasons, we model the Tree Ring widths time series using long-memory models; specifically, the ARFIMA models are used to estimate, diagnose, and compare forecasts of the number of tree rings for upcoming years.

In order to illustrate the usage of package functions, we consider a fitted ARFIMA$(1,d,1)$ model. For this  model, we have implemented the Whittle algorithm and computed the exact variance-covariance matrix to compare with the Hessian method. Afterward, we compare the Sowell method for computing the ACVF function with the sample ACVF. Other functions have also been implemented and illustrated in this section.

\subsection[The implementation of the Whittle algorithm]{The implementation of the Whittle algorithm}

The estimation of the fractionally, autoregressive, and moving-average parameters has been studied
by several authors (\citet{lr:07}; \citet{b:09}; \citet{q:11}). A widely used method is the approximate MLE method of \citet{b:09}.
In our study, estimation of the ARFIMA$(p,d,q)$ model using the corresponding Whittle method \citep{bn:21} is described in Section 2.1
and this model is fitted by using the {\tt arfima.whittle} function. To apply the Whittle algorithm to the  {TreeRing} time series
as an example, we use the following command considering an ARFIMA$(1,d,1)$ model.\\

\begin{verbatim}
R> y <- data(TreeRing)
R> model <- arfima.whittle(series = y, nar = 1, nma = 1, fixed = NA)
\end{verbatim}

Note that the option {\tt fixed} (for fixing parameters to a constant value) has been implemented. This option allows the user  to fix the parameters $d$, $\phi_1$, or $\theta_1$, in order of occurrence. For example, in our ARFIMA$(1,d,1)$ model, we can set the parameter $d$ to be equal to zero. Consequently, we obtain the estimation of a simple ARMA$(1,1)$ model. The object {\tt model} is of class {\tt arfima} and provides the following features:

\begin{itemize}
\item estimation of $d$ and ARMA parameters;
\item standard deviation errors obtained by the Hessian method and the respective  {\tt t value} and  {\tt Pr(>|t|)} terms;
\item the log-likelihood function performed in the {\tt arfima.whittle.loglik()} function;
\item innovation standard deviation estimated by the ratio of the theoretical spectrum;
\item residuals from the fitted model.
\end{itemize}

The commands  {\tt plot()},  {\tt residuals()},  {\tt summary()},  {\tt tsdiag()} and  {\tt print()}  have been adapted to this model class of S3 method. The  {\tt summary()} option shows the estimated parameters, the Hessian standard deviations, the $t$-statistic, and their respectively $p$-values. The computation of the long-memory parameter $d$ as well as the autoregressive $\{\phi_1,...,\phi_p\}$ and moving average $\{\theta_1,...,\theta_q\}$ parameters can be handled quickly for moderate sample sizes. Printing the {\tt model} object by the  {\tt summary()} function shows the items mentioned before as

\begin{verbatim}
R> summary(model)
$call
arfima.whittle(series = y, nar = 1, nma = 1)

$coefmat
          Estimate  Std. Error   t value   Pr(>|t|)
d        0.1058021  0.04813552  2.198004 0.02794879
phi 1    0.3965915  0.03477914 11.403142 0.00000000
theta 1 -0.2848590  0.03189745 -8.930462 0.00000000

$sd.innov
[1] 35.07299

$method
[1] "Whittle"

attr(,"class")
[1] "summary.arfima"
\end{verbatim}

Furthermore, we evaluate the Whittle estimation method for long-memory models by using the
theoretical parameters versus the estimated parameters (see Figure~\ref{sim}). In addition, we compare the
exact standards deviations of the parameters from Section~2.2 with those
obtained by the Hessian method. These results and the evaluation of the Whittle estimations are illustrated in Table~\ref{models}.

\begin{table}
\caption{Summary of estimated parameters for several ARFIMA models.}\label{models}
\begin{center}
\begin{scriptsize}
\begin{tabular}{ccccccccccccc}
  \hline
    & \multicolumn{3}{c}{FN$(d)$} & \multicolumn{3}{c}{ARFIMA$(1,d,0)$} & \multicolumn{3}{c}{ARFIMA$(0,d,1)$} & \multicolumn{3}{c}{ARFIMA$(1,d,1)$ }\\
  \cline{2-4} \cline{5-7} \cline{8-10} \cline{11-13}
  Par. & Est & Hessian & Exact & Est & Hessian & Exact & Est & Hessian & Exact & Est & Hessian & Exact\\
  \hline
  $d$ & 0.195 & 0.048 & 0.023 & 0.146 & 0.048 & 0.038 & 0.156  & 0.048 & 0.035 & 0.106 & 0.048 & 0.063 \\
  $\phi$ & - & - & - & 0.072 & 0.029 & 0.049 & - & - & - & 0.397 & 0.035 & 0.282\\
  $\theta$ & - & - & - & - & - & - & 0.059 & 0.029 & 0.045 & -0.285 & 0.032 & 0.254\\
  \hline
\end{tabular}
\end{scriptsize}
\end{center}
\end{table}

\subsection[The implementation of exact variance-covariance matrix]{The implementation of exact variance-covariance matrix}

The {\tt var.afm()} function shows the exact variance-covariance matrix and the standard
deviations. The computation of the integrals of the expression \eqref{gamma} is carried out by
using the Quadpack numeric integration method \citep{j:17} implemented in the  {\tt integrate()}
function ({\tt stats} package). Note that the functions involved in these integrals diverge in the
interval $\lambda=[-\pi,\pi]$. However, they are even functions with respect to $\lambda$. Thus, we
integrate over $[0,\pi]$ and then multiply the result by two. Now, by using the central limit
theorem discussed in Section 2.2, we can obtain the asymptotic approximation of the estimated
parameters standards errors
$\mbox{SE}(\widehat{{\bf\Omega}})_i=(\frac{1}{n}[\widehat{{\bf\Sigma}}^{-1}]_{ii})^{1/2}$
of an ARFIMA model, where $\widehat{{\bf\Omega}}=(\widehat{d},\widehat{\phi}_1,\ldots,\widehat{\phi}_p,\widehat{\theta}_1,\ldots,\widehat{\theta}_q)$ and $[\widehat{{\bf\Sigma}}^{-1}]_{ii}$ corresponds to the $i$th diagonal components of the matrix $\widehat{{\bf\Sigma}}^{-1}$ for $i=\{1,...,p+q+1\}$.

\begin{table}
\caption{Akaike's criterion for $\widehat{{\bf\Omega}}=(\widehat{d},\widehat{\phi}_1,\widehat{\phi}_2,\widehat{\theta}_1,\widehat{\theta}_2)$ with $p$-values obtained for the Hessian standard deviation.}\label{aic}
\begin{center}
\begin{tabular}{ccccc}
  \hline
  $p$ & $q$ & AIC & $\widehat{d}$ & p-value\\
  \hline
   0 &  0 & \bf{-37.018} &  0.196 &  0\\
   0 &  1 & \bf{-35.008} &  0.156 &  0.001\\
   0 &  2 & -33.000 &  0.113 &  0.018\\
   1 &  0 & \bf{-35.007} &  0.146 &  0.002\\
   1 &  1 & \bf{-33.004} &  0.106 &  0.028\\
   1 &  2 & -30.996 &  0.142 &  0.003\\
   2 &  0 & -33.002 &  0.111 &  0.021\\
   2 &  1 & -30.995 &  0.130 &  0.007\\
   2 &  2 & -28.915 &  0.191 &  0\\
   \hline
\end{tabular}
\end{center}
\end{table}

By using the Whittle estimators, we search for the lowest AIC \citep[Akaike Information Criterion,][]{aa:01} given by
$\mbox{AIC}(\widehat{\omega})=-2[\log\,L(\widehat{\omega})-(p+q+1)]$ over a class of ARFIMA models with $p,q\in \{0,1,2\}$, where
$\widehat{\omega}$ is a subset of $\widehat{{\bf\Omega}}$ and $L(\widehat{\omega})$ is the likelihood associated with $\widehat{\omega}$.
From Table~\ref{aic}, we can see that the fractionally differenced model ARFIMA($0,d,0$) has the lowest AIC. Candidate models are marked
in bold in Table~\ref{aic}.

\begin{figure}[h!]
    \centering
    \includegraphics[width=10cm,height=8cm]{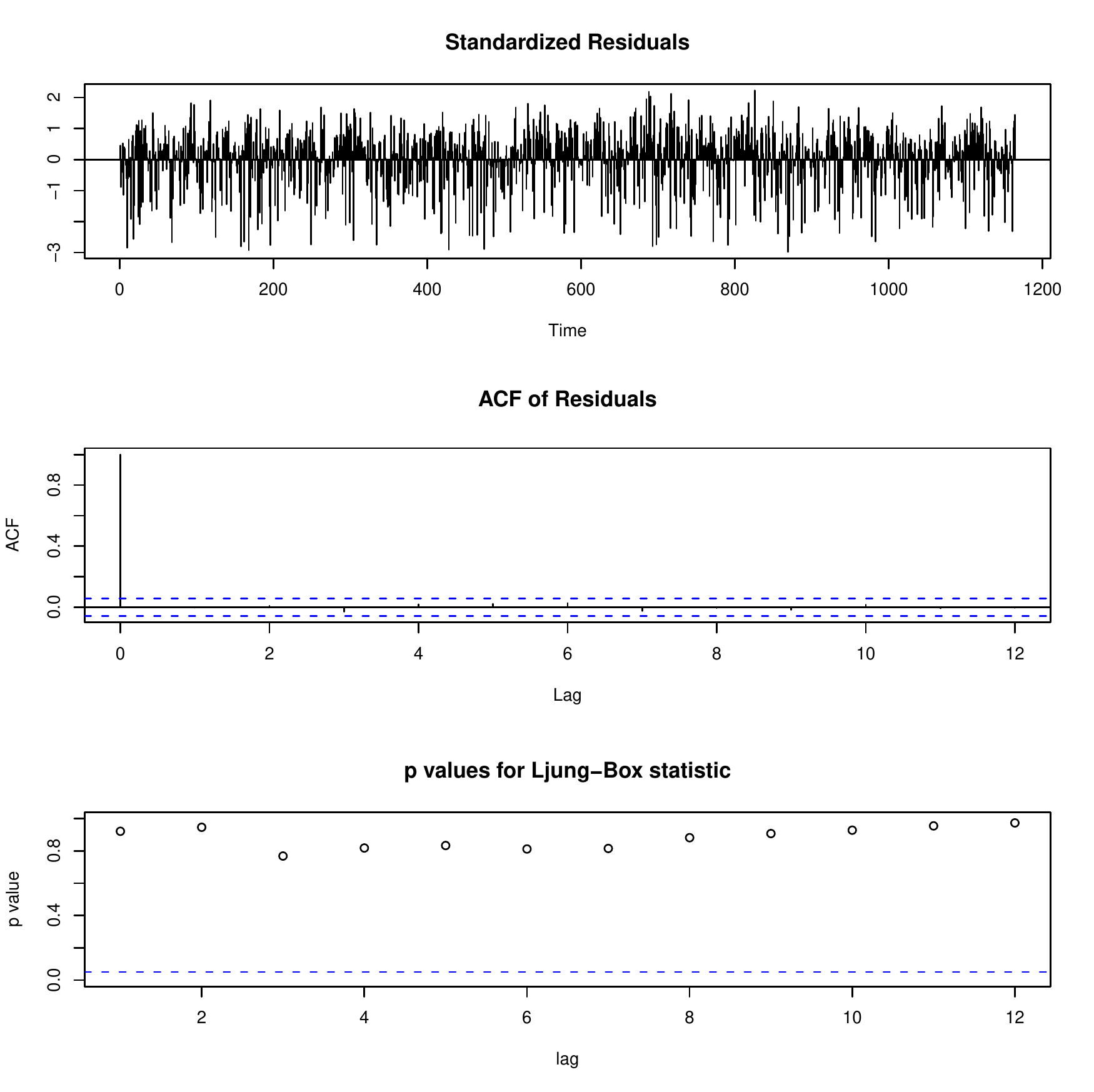}
\caption{Plots of residuals analysis from  {tsdiag} standard command adapted to ARFIMA model residuals.}\label{units}
\end{figure}

Additionally, we propose a technique for obtaining the spectral density associated with ARFIMA and ARMA processes in  {\tt spectrum.arfima()} and {\tt spectrum.arma()}, respectively. This is done by using the {\tt polyroot()} function of the {\tt polynom} package to compute the roots of the polynomials $\Phi(e^{-i\lambda})$ and $\Theta(e^{-i\lambda})$. Both functions need the estimated ARFIMA parameters and the  {sd.innov} innovation standard estimation given by an object of {\tt arfima} class. For the spectrum density and periodogram, see Section~2 and Subsection~2.1, respectively. Since the calculation of the FFT has a numerical complexity of the order $O[n\log_2(n)]$, this approach produces a very fast algorithm to estimate the parameters. It is possible to obtain the FFT through the {\tt fft()} function based on the method proposed by \citet{f:19}.

\begin{figure}[h!]
    \centering
    \includegraphics[width=9cm,height=10cm]{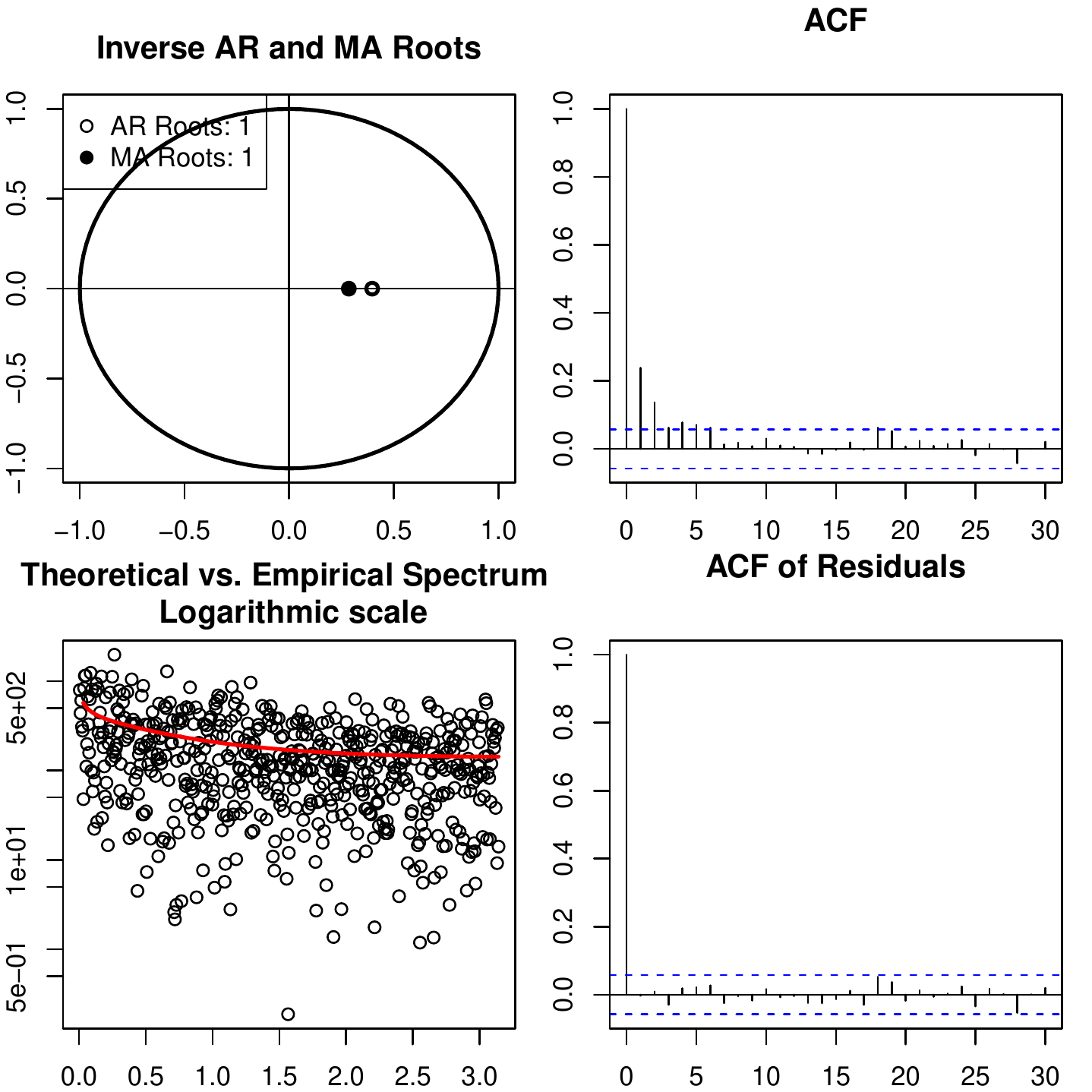}
\caption{Diagnostic Plots made by the {\tt plot()} command. Left: Plots of unit roots circle along with the root provided by AR and MA polynomials and Theoretical (red line) vs. Empirical Spectrum (back points) plot. Right: ACF plots of Tree Rings and ARFIMA$(1,d,1)$ model residuals.}\label{plot}
\end{figure}

\subsection[The implementation of the diagnostic functions]{The implementation of the diagnostic functions}

We have also implemented a very practical function called  {\tt check.parameters.arfima()}. This verifies whether
the long-memory parameter $d$ belongs to the interval $(-1,0\mbox{.}5)$ and whether the roots of the fitted ARMA
parameters lie outside the unit disk. This function was incorporated in the {\tt plot()} command. In the first plot
of Figure~\ref{plot}, we can see that the roots of the AR and MA polynomials lie outside the unit disk, according
to the assumptions of stationarity solutions of (\ref{arfima}) presented in Theorem~\ref{T1} (see Section~2). Alternatively, the
{\tt check.parameters.arfima()} that takes an {\tt arfima}-class object, gives {\tt TRUE/FALSE}-type
results indicating whether the parameters pass the requirement for a stationary process.

\begin{figure}[h!]
    \centering
    \includegraphics[width=6cm,height=5cm]{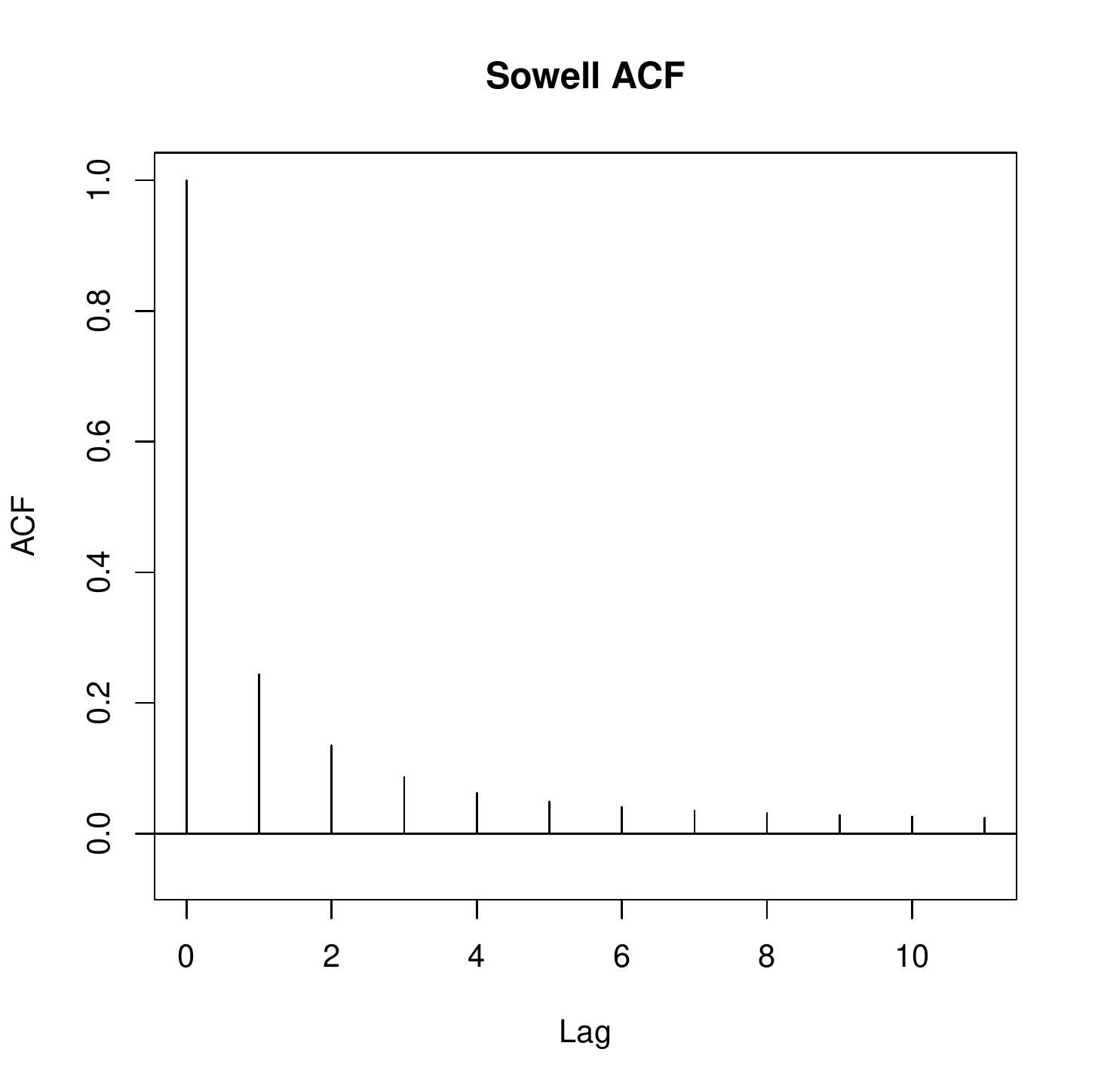}
    \includegraphics[width=6cm,height=5cm]{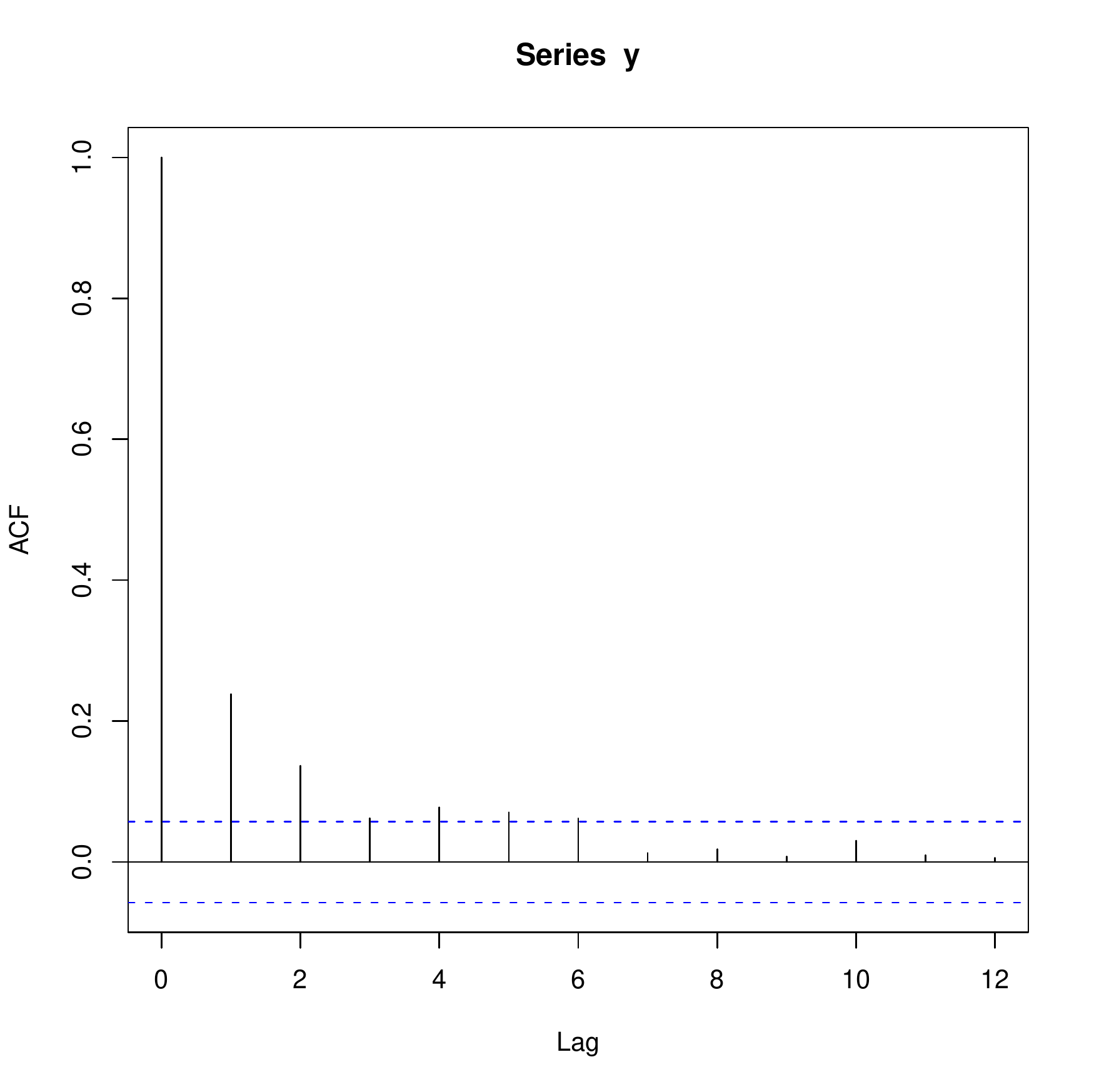}
\caption{Theoretical (left) and Empirical (right) ACF of selected ARFIMA model.}\label{cov}
\end{figure}

\begin{figure}[h!]
    \centering
    \includegraphics[width=12cm,height=13cm]{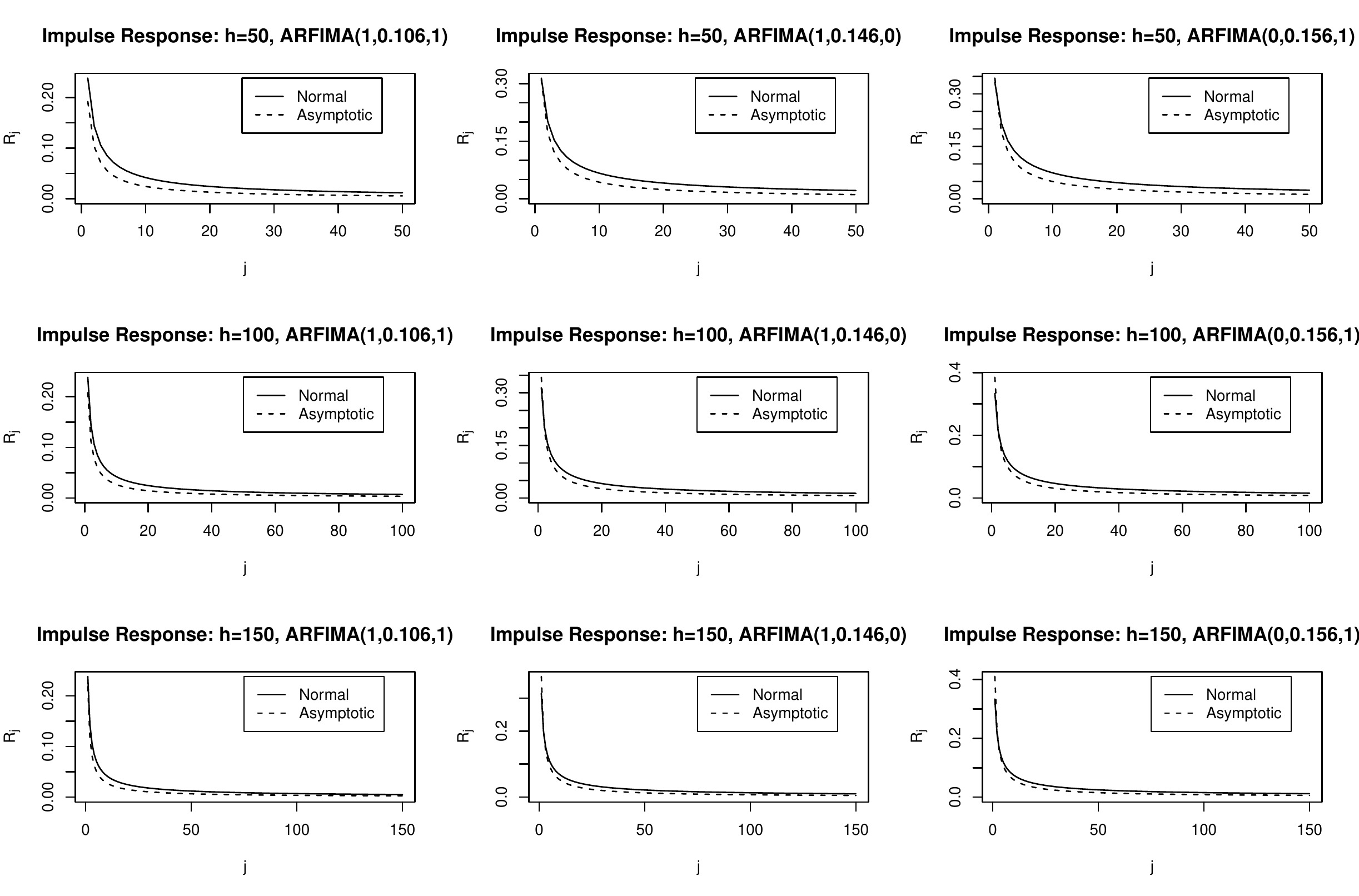}
\caption{Plots of IRFs methods for $h=50$, 100 and 150 lags.}\label{IR}
\end{figure}

Additionally, an adaptation of the function  {\tt tsdiag()} can be found in this package.
This is implemented in an S3-type  {\tt arfima} class method and shows three plots for analyzing the
behavior of the residual from the fitted ARFIMA model. This function has  additional arguments such
as the number of observations  {n} for the standardized residuals and critical $p$-value
{\tt alpha}. Figure~\ref{units} illustrates these results, where, the residuals are white
noise at a confidence level of $\alpha=95\%$.

On the other hand, the $R_j$ (IRF) illustrated in Figure~\ref{IR} decays exponentially fast, at
a rate of $j^{d-1}$ because, these functions inherit the behavior of $\eta_j$. This behavior is typical for ARFIMA
models, as reported by \citet{ha:12}, \citet{w:12} and \citet{er:10}. Figure~\ref{IR} shows some $R_j$ curves
associated with the three models considered in Table~\ref{models} for the asymptotic method by formula (\ref{IR2})
(labeled  {\tt Asymptotic} in the plot) and the counterpart method by formula (\ref{IR1}) (labeled  {\tt Normal} in the plot).
Note that for a large value of $j\approx50$, both methods tend to converge, and, the curves make an inflexion in the value $j\approx10$.
Note that the  {\tt Asymptotic} approximation tends to be equal to the {\tt Normal} method in the measure that the input  {h} lag increases
(see plots for  {h=150} in Figure~\ref{IR}). These IRFs are available in the function  {\tt ir.arfima()}, with arguments  {\tt h} to evaluate the IRFs
over a particular $h$ lag and,  {model} for an object  {\tt arfima.whittle}.
The {\tt ir.arfima()} function produces the vectors  {\tt RE} and  {\tt RA} for  {\tt Normal} and  {\tt Asymptotic} IRFs.

The exact Sowell autocovariance computation obtained by {\tt rho.sowell()} and the sample
autocorrelation obtained by the  {\tt ACF()} command are applied to tree ring time series.
In Figure~\ref{cov}, the blue dotted lines in the first plot correspond to the $\{\mp
2/\sqrt{n}\}$ significance level for the autocorrelations. The function  {\tt rho.sowell()} requires
the specification of an object of  class  {\tt arfima} in the  {object} option that, by
default, is {\tt NULL}. But, if  {\tt object=NULL}, the user can incorporate the ARFIMA
parameters and the innovation standard deviation. Alternatively, the implemented
{\tt plot} option gives a graphical result similar to the  {\tt ACF()} command in the sample
autocorrelation. We can see the similarity of both results for the discussed model. The ACVF
implementation is immediate but, for the calculation of the Gaussian hypergeometric functions, we
use the {\tt hypergeo()} function from the {\tt hypergeo} package. For values of $h>50$, we use
the approximation \eqref{ACVFa} reducing considerably the computation time as compared to the Sowell
algorithm.

On the other hand, the {\tt rho.sowell()} function is required by the function {\tt smv.afm()}.
The function {\tt smv.afm()} calculates the variance of the sample mean of an
ARFIMA process. When the argument {\tt comp} is {\tt TRUE}, the finite sample variance is calculated,
and when {\tt comp} is {\tt FALSE}, the asymptotic variance is calculated.
%

\begin{figure}[h!]
    \centering
    \includegraphics[width=5.1cm,height=5.5cm]{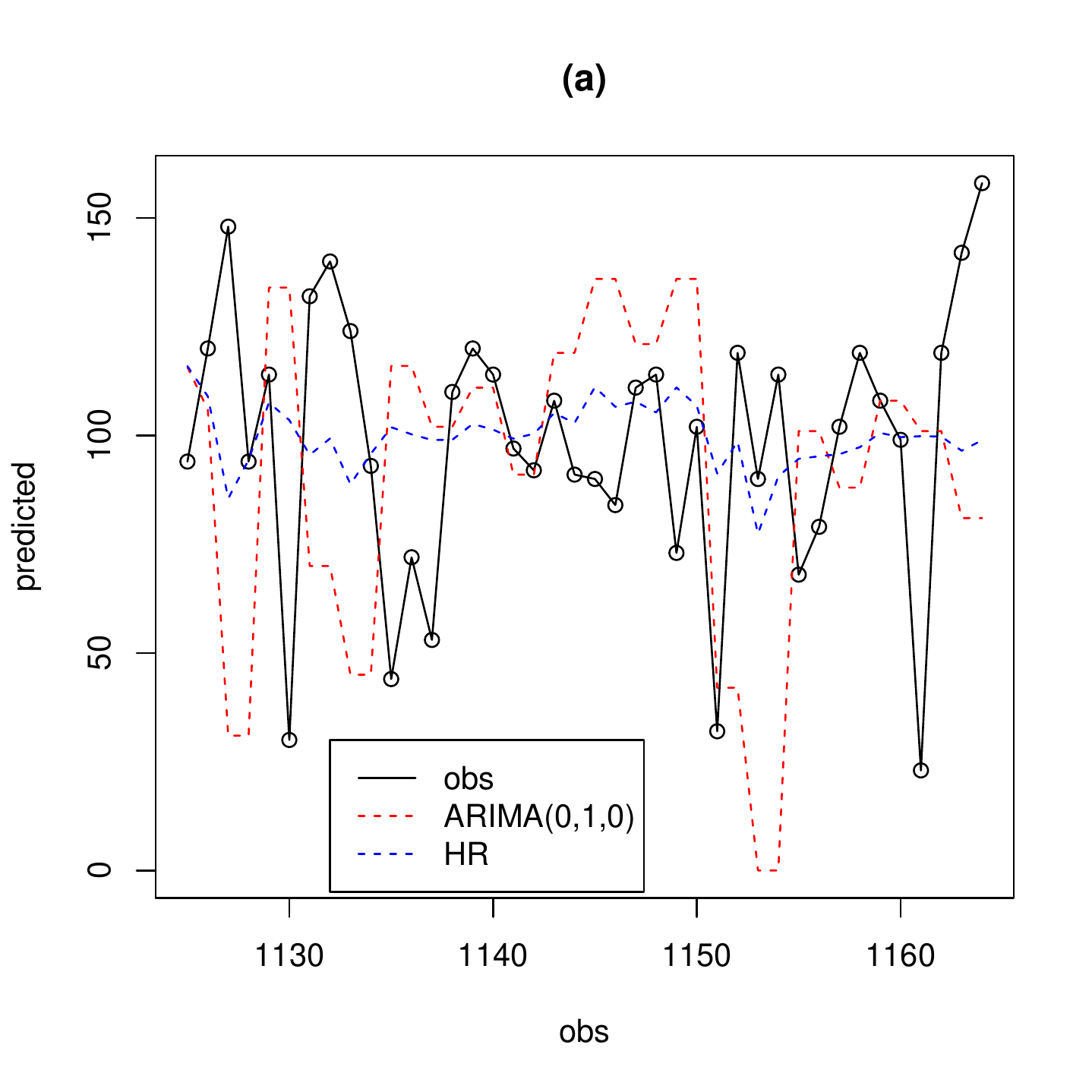}
    \includegraphics[width=5.1cm,height=5.5cm]{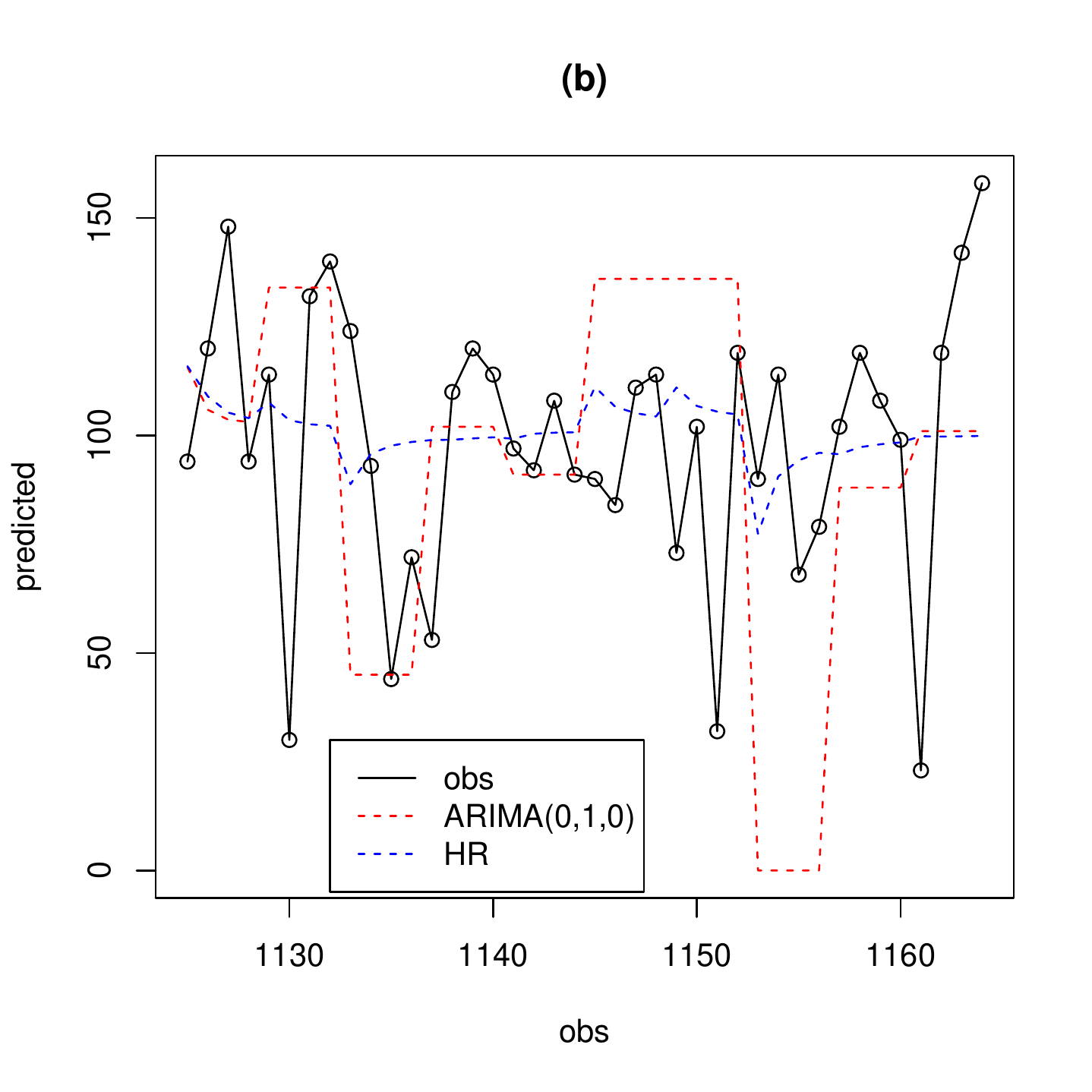}
    \includegraphics[width=5.1cm,height=5.5cm]{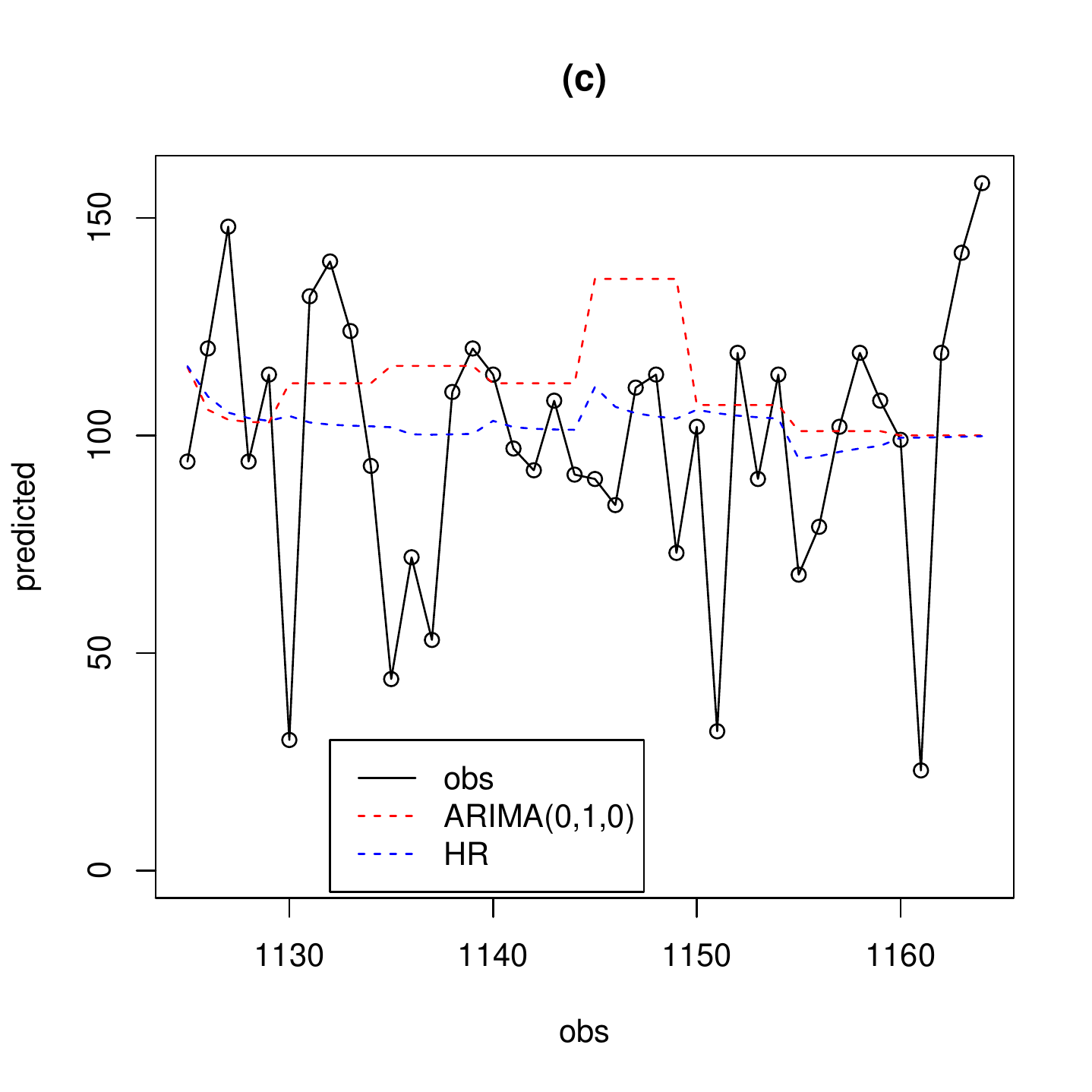}
\caption{Plots of out-of-sample predictions for (a) $\tau=2$, (b) $\tau=4$ and (c) $\tau=5$ for two models: ARIMA(0,1,0) in red and ARFIMA$(1,d,1)$ in blue with Haslett \& Raftery estimator (HR).}\label{os}
\end{figure}

\begin{table}[h!]
\begin{center}
\begin{scriptsize}
\caption{Summary of p-values of the GW test for each HAC estimator, $\tau=\{2,4,5\}$ prediction horizon parameters and for the estimator
methods B (Benchmark model), ML (ARFIMA models using Maximum Likelihood estimator) and HR (ARFIMA models using Haslett \& Raftery
estimator) over 40 observations of samples from the {\tt TreeRing} data set. The $p$-values marked in {\bf bold} are lower than the probability
(0.05) related to a 5\% confidence level.} \label{gwt}\vspace{0.5cm}
\begin{tabular}{|c|cccc|cccc|cccc|}
  \hline
  HAC  &  \multicolumn{12}{|c|}{Prediction Horizon Parameter} \\
  Estimators &  \multicolumn{4}{|c}{$\tau=2$} & \multicolumn{4}{c}{$\tau=4$} & \multicolumn{4}{c|}{$\tau=5$} \\
   \hline
        &    & $B$ & $ML$  & $HR$    &    & $B$ & $ML$  & $HR$    &    & $B$ & $ML$ & $HR$  \\
        & $B$  & -  &\bf 0.001    &\bf  0.001     & $B$  & -  &\bf  0.031   &\bf  0.031     & $B$  & -  & 0.126   & 0.126\\
  HAC   & $ML$ &\bf 0.001  &  -   &  0.307     & $ML$ &\bf 0.031  &  -   &  0.559     & $ML$ & 0.126  &  -  & 0.689\\
        & $HR$ &\bf 0.001  &  0.307   &   -    & $HR$ &\bf 0.031  &  0.559   &    -   & $HR$ & 0.126  & 0.689   & -\\
   \hline
            &    & $B$ & $ML$ & $HR$ &    & $B$ & $ML$ & $HR$ &    & $B$ & $ML$ & $HR$  \\
  Newey \&  & $B$  & -  &\bf  0.002  &\bf 0.002   & $B$  & -  &\bf 0.051   &\bf 0.051   & $B$  & -  & 0.174   & 0.174\\
  West      & $ML$ &\bf 0.002  &  -  &\bf  0.005  & $ML$ &\bf 0.051  &  -  & 0.281   & $ML$ & 0.174  &  -  & 0.479\\
            & $HR$ &\bf 0.002  &\bf 0.005   &  -  & $HR$ &\bf 0.051  &  0.281  &  -  & $HR$ & 0.174  & 0.479   & -\\
   \hline
            &    & $B$ & $ML$ & $HR$ &    & $B$ & $ML$ & $HR$ &    & $B$ & $ML$ & $HR$  \\
  Lumley \& & $B$  & -  &\bf  0.001  &\bf  0.001  & $B$  & -  & \bf 0.023   &\bf 0.023  & $B$  & -  &  0.096  & 0.096\\
  Heagerty  & $ML$ &\bf 0.001  &  -  & 0.486   & $ML$ &\bf 0.023  &  -   & 0.652  & $ML$ & 0.096  &  -  & 0.762\\
            & $HR$ &\bf 0.001  &  0.486  &  -  & $HR$ &\bf 0.023  &  0.652   &  - & $HR$ & 0.096  &  0.762  & -\\
   \hline
            &    & $B$ & $ML$ & $HR$ &    & $B$ & $ML$ & $HR$ &    & $B$ & $ML$ & $HR$  \\
           & $B$  & -  &\bf 0.003   &\bf 0.003   & $B$  & -  &\bf  0.041  &\bf 0.041   & $B$  & -  &  0.206  & 0.206\\
  Andrews  & $ML$ &\bf 0.003  &  -  &\bf  0.018  & $ML$ &\bf 0.041  &  -  &  0.283  & $ML$ & 0.206  &  -  & 0.456\\
           & $HR$ &\bf 0.003  &\bf  0.018  &  -  & $HR$ &\bf 0.041  &  0.283  &  -  & $HR$ & 0.206  & 0.456  & -\\
  \hline
\end{tabular}
\end{scriptsize}
\end{center}
\end{table}

\subsection[Forecasting evaluations]{Forecasting evaluations}

The GW method implemented in  {\tt gw.test()} for evaluating forecasts proposed by \citet{m:06} compares two vectors of
predictions,  {\tt x} and  {\tt y}, provided by two time series models and a data set  {\tt p}. We
consider that it is relevant to implement this test to determine if the predictions produced by a
time series model (e.g., ARFIMA) process good forecasting qualities. This \emph{test for
predictive ability} is of particular interest since it considers
the  {\tt tau} prediction horizon parameter or  {\tt ahead} in the case of  {\tt pred.arfima()} function.
Alternative methods are discussed, for instance, by \citet{llt:05}. If  {\tt tau=1}, the
{\em standard statistic simple regression estimator} method is used. Otherwise, for values of
 {\tt tau} larger than 1, the  method chosen by the user is used in the  {\tt method} option. The
available methods for selection are described below. They include several Matrix Covariance Estimation
methods but, by default, the HAC estimator is used in the test. The user can select between
the several estimators of the  {\tt sandwich} package mentioned before:

\begin{itemize}
\item  {\tt HAC}: Heteroscedasticity and Autocorrelation Consistent (HAC) Covariance Matrix Estimation by  {\tt vcovHAC()} function (\citet{we:23};
\citet{tr:24}).
\item  {\tt NeweyWest}: Newey-West HAC Covariance Matrix Estimation by  {\tt NeweyWest()} function (\citet{z:14}).
\item  {\tt LumleyHeagerty}: Weighted Empirical Adaptive Variance Estimation by  {\tt weave()} function (\citet{a:13}).
\item  {\tt Andrews}: Kernel-based HAC Covariance Matrix Estimation by  {\tt kernHAC()} function (\citet{bs:01}; \citet{cf:02}).
\end{itemize}

This test gives the usual results of the general test implemented in such as the GW
statistic in  {\tt statistic}, the alternative hypothesis in   {\tt alternative}, the $p$-value
in  {\tt p.value}, others such as the method mentioned before in  {\tt method}, and the name of
the data in  {\tt data.name}. In some studies, the GW test is used to compare selected models versus benchmark models such as ARMA, ARIMA, or SARIMA models
\citep[e.g.][]{jkb:04}. So, to illustrate the GW test performance, we simulate the out-of-sample prediction exercise
through mobile windows  for  {\tt TreeRing} data sets considering the first 1124 observations and, later, forecasting
40 observations using three models: the ARIMA(0,1,0) benchmark model, the ARFIMA$(p,d,q)$ with MLE estimator, and the
Haslett \& Raftery estimator (HR) using the algorithm of the automatic  {\tt forecast()} function implemented in the  {\tt forecast} package
by \citet{q:11} algorithm. In Figure~\ref{os}, the GW test compares the out-of-sample predictions of the ARIMA(0,1,0) and ARFIMA(p,d,q) model.
It is important to note that the goal of this test is only to compare prediction abilities between models.

Finally, we study a more general simulation, comparing the three predictors vectors with 40 real observations using
 {\tt gw.test()} function considering the hypotheses testing {\tt alternative="two.sided"} to contrast significant
differences between predictions. In addition, we consider the four HAC estimators mentioned in the beginning of
this subsection and prediction horizon parameters $\tau=\{2,4,5\}$.
The results are summarized in Table~\ref{gwt} and Figure~\ref{os}. We can see that the differences in the prediction ability
between B vs. MLE and B vs. HR are significant for $\tau=2$ and 4 but, between MLE and HR they are not unequal for the
three considered values of $\tau$. Given the non-significance of the MLE-HR test $p$-value, the MLE model is not considered in
Figure~\ref{os}. As expected, the increments of the prediction horizon parameter showed that the different prediction
abilities of the models tend to be zero because the time series uncertainty tends to increase. Consequently, the individual prediction performance
of each model is not considered by the test.

\section{Conclusions}

We developed the {\tt afmtools} package with the goal of incrementing the necessary utilities
to analyze the ARFIMA models and, consequently, it is possible to execute several useful commands
already implemented in the   long-memory packages. In addition, we have provided the
theoretical results of the Whittle estimator, which were applied to the Tree Ring data base.
Furthermore, we have performed a brief simulation study for evaluating
the estimation method used herein and also have evaluated the properties of its log-likelihood function.
The numerical examples shown here illustrate the different capabilities and features of the
{\tt afmtools} package; specifically, the estimation, diagnostic, and forecasting functions.
The {\tt afmtools} package would be improved by incorporating other functions related to
change-point models and tests of unit roots, as well as other
important features of the models related to long-memory time series.\\

\noindent {\large\bf Acknowledgment}
Wilfredo Palma would like to thank the support from Fondecyt Grant 1120758.\\

%
%

\end{document}